\documentclass[structabstract]{aa}  

\usepackage{txfonts,xcolor}
\usepackage{graphicx}
\usepackage{booktabs}
\usepackage{siunitx}
\usepackage{colortbl}
\usepackage[pdftex, colorlinks=true, linkcolor=blue, citecolor=blue, urlcolor=blue]{hyperref}
\usepackage{nameref} 

\def\deg{$^{\circ}~$}

\newcommand\rev[1]{#1}
\newcommand\final[1]{#1}
\begin{document}

   \title{The grazing angle icy protoplanetary disk PDS 453}

    \author{L. Martinien\inst{1}
   \and F. M\'enard \inst{1}
   \and G. Duch\^ene\inst{1,2}
   \and R. Tazaki\inst{1,11}
   \and M.D. Perrin\inst{3}
   \and K.R. Stapelfeldt\inst{4}
   \and C. Pinte\inst{5}
   \and S.G. Wolff\inst{6}
   \and C. Grady\inst{7}
   \and C. Dominik\inst{8}
   \and M. Roumesy\inst{1}
   \and J. Ma\inst{1}
   \and C. Ginski\inst{9}
   \and M. Benisty\inst{10,1}
    \and D.C. Hines\inst{3}
   \and G. Schneider\inst{6}
   }
   \institute{Univ. Grenoble Alpes, CNRS, IPAG, 38000 Grenoble, France \\
    \email{laurine.martinien@univ-grenoble-alpes.fr}
    \and Astronomy Department, University of California Berkeley, Berkeley CA 94720-3411, USA
    \and Space Telescope Science Institute, Baltimore, MD 21218, USA
    \and Jet Propulsion Laboratory, California Institute of Technology, 4800 Oak Grove Drive, Pasadena, CA 91109, USA
    \and Monash Centre for Astrophysics (MoCA) and School of Physics and Astronomy, Monash University, Clayton, Vic 3800, Australia
    \and Steward Observatory and the Department of Astronomy, The University of Arizona, 933 N Cherry Ave, Tucson, AZ, 85719, USA
    \and Eureka Scientific, 2452 Delmer Street, Suite 1, Oakland, CA 96402, USA
    \and Anton Pannekoek Institute for Astronomy, University of Amsterdam, Science Park 904, 1098 XH Amsterdam, The Netherlands
    \and Centre for Astronomy, Dept. of Physics, National University of Ireland Galway, University Road, Galway H91 TK33, Ireland
    \and Universit\'e C\^ote d’Azur, Observatoire de la C\^ote d’Azur, CNRS, Laboratoire Lagrange, F-06304 Nice, France
    \and Department of Earth Science and Astronomy, The University of Tokyo, Tokyo 153-8902, Japan
      }

   \date{Received 12 July 2024; accepted 05 November 2024}

  \abstract
   {Observations of highly inclined protoplanetary disks provide a different point of view, in particular a more direct access to their vertical structure when compared to less inclined, more pole-on disks. 
  }
   {PDS 453 is a rare highly inclined disk where the stellar photosphere is seen at grazing incidence on the disk surface. Our goal is take advantage of this geometry to  constrain the structure and \rev{composition} of this protoplanetary disk, in particular the fact that it shows a 3.1 $\mu$m water ice \rev{band} in absorption that can be related uniquely to the disk.}
   {We observed the system in polarized intensity with the
   VLT/SPHERE instrument, as well as in polarized light and total intensity using the HST/NICMOS camera. Infrared archival \rev{photometry and a spectrum showing the water ice band are} used to model the spectral energy distribution \final{under Mie scattering theory.} 
   \rev{Based on these data, we fit a model using} the radiative transfer code MCFOST to retrieve the geometry and dust and ice content of the disk.}
  {PDS 453 has 
  the typical morphology of a highly inclined system with two reflection nebulae where the disk partially attenuates the stellar light. The upper nebula is brighter than the lower nebula and shows a curved surface brightness profile in polarised intensity, indicating a ring-like structure. With an inclination of 80° estimated from models, the line-of-sight crosses the disk surface and a combination of absorption and scattering by ice-rich dust grains produces the water ice band.  

}
   {PDS 453 is seen at high inclination and is composed of a mixture of silicate dust and water ice. The radial structure of the disk includes a significant jump in density and scale height at a radius of 70 au in order to produce a ring-like image. The depth of the 3.1\,$\mu$m \rev{water ice} band depends on the amount of water ice, until it saturates when the optical thickness along the line-of-sight becomes too large. Therefore, quantifying the exact amount of water from absorption bands in edge-on disks requires a detailed analysis of the disk structure and tailored radiative transfer modeling. Further observations with JWST and ALMA will allow to refine our understanding of the structure and content of this interesting system.
   }

   \keywords{ 
   protoplanetary disks -- Stars: Individual: PDS 453 -- Stars: variable: Herbig stars
   }

  \maketitle
%
\section{Introduction}

Protoplanetary disks around young stars are the birth places of planets \rev{\citep[e.g.,][]{Keppler2018}}. High contrast and high angular resolution images with modern facilities unambiguously revealed a wide variety of structures within disks, e.g., gaps, rings, and spiral arms, each of these hinting at the presence of undetected planets. These structures are detected both in scattered light in the optical and NIR, e.g., with HST and VLT/SPHERE \citep[for a recent review, see][]{benisty2023}, and in thermal continuum emission at longer \rev{millimeter} wavelengths, e.g., with ALMA \citep{Andrews2018, Long2018}. 
To make progress in that direction, it is necessary to estimate the structural parameters of protoplanetary disks in more details.

Interest in a particular sub-group of disks, those seen at high inclinations, has increased in recent years \citep[e.g.,][]{Burrows1996, Watson2007, Duchene2024, Villenave2024}
as they offer a different view, revealing more directly the vertical distribution of material above and below the disk midplane \rev{\citep{Andrews2018, Long2018}}. This is complementary to other studies probing more directly the radial distribution of material, more easily seen in star+disk systems at intermediate inclinations or close to pole-on. In particular, edge-on disk images have allowed the direct identification of vertical dust settling. A general trend is emerging from the sample observed so far that small dust particles (traced by optical/near-infrared scattered light) are well coupled to the gas, producing two reflection nebulae well separated from the disk midplane. On the contrary, the millimeter dust thermal emission (tracing larger mm-sized particles) is massively concentrated in a vertically very thin layer \citep{villenave2020}. 

Edge-on disks also offer the possibility to probe absorption \rev{band} of solid state material, e.g., H$_2$O, CO, and CO$_2$ ice coatings on grains \rev{\citep{Terada2017, Sturm2023, Sturm_2024}}. This is relevant in the context of grain growth because ice coatings on grains are expected to favor sticking, and therefore growth \citep[e.g.,][]{Johansen_2014}. Such studies are possible when the line-of-sight to the photosphere is at grazing incidence on the disk surface \citep{Glauser2008}. New observations with JWST are primed to yield new insight on this topic, as recently demonstrated by \citet{Sturm2023} who attempted to estimate the column density of different species of ices in the HH\,48 system. Interestingly, PDS 453 is another system where this can be performed \rev{to test whether the behavior of absorption ice bands is ubiquitous in these disks}.

PDS 453 is an F2e-type star located at the periphery of the Scorpius-Centaurus OB association \citep{Sartori_2010}. Assuming PDS 453 belongs to the association and in the absence of a reliable Gaia parallax due to the spatially extended nature of the source, it is assumed to be a young \citep[$\approx5$--10\,Myr][]{preibisch2008nearest, ratzenbock23} intermediate-mass star located at 130\,pc. Given its spectral type, it is intermediate between a low-mass T\,Tauri star and a Herbig Ae star \citep{Vioque_2018}. The disk of PDS 453 was first identified by \citet{Perrin2006PhDT.......203P} using Lick Observatory's Shane 3-m telescope with the IRCAL polarimeter. A finer image was subsequently obtained with NICMOS onboard the {\it Hubble Space Telescope} revealing the nearly edge-on nature of the disk \citep{2010AAS...21542812P}. \citet{Terada_2017} presented a \rev{3 $\mu$m} spectra acquired with the Infrared Camera and Spectrograph (IRCS) instrument at the Subaru telescope revealing a water ice absorption band at 3.1 $\mu$m \rev{as well as two NIR images acquired by the Subaru AO system showing a bright central point source.} 

In this paper we present \rev{archival} HST/NICMOS and \rev{new} VLT/SPHERE \rev{data, providing the sharpest and highest contrast image of the disk to date.}
We also present a radiative transfer model of the system to match the HST/NICMOS and VLT/SPHERE images, the HST/NICMOS polarimetry, the water ice band absorption \rev{detected with Subaru/IRCS}, and the spectral energy distribution. The data and results are presented in Section~\ref{sec:ObsRes}. In Section~\ref{sec:modeling} we describe the radiative transfer model. The results from the modeling are presented in Section~\ref{sec:results} and they are discussed in Section~\ref{sec:discussion}. Section~\ref{sec:conclusion} summarizes the main conclusions. 

\section{Data and Observational Results}
\label{sec:ObsRes}


\begin{table*}[h!]
	\centering
 \caption{Observation log.} 
	\label{tab:telescopes}

        \footnotesize 
 \begin{tabular}{ccccc}
		 \toprule
          \toprule
		Observation & Instrument & \rev{Filters ($\lambda_{c}\tablefootmark{a}$ ($\mu$m))} & \rev{$t_{exp}$} \tablefootmark{b} (s) &  
            Observation date \\
		 \midrule
		  PI image & VLT/SPHERE & \rev{BB$\_$H (1.6)} & 2304  & 2018/06/22 \\
		I image & HST/NICMOS & \rev{F110W (1.1)} 	& 192     & 
            2008/05/28\\
		P image & HST/NICMOS & \rev{POL*L (2.0)}  	& 512  & 2008/05/28\\
       
		\bottomrule
     \label{tab:observations}
	\end{tabular}
\tablefoot{
\tablefoottext{a}{Central Wavelength.}
\tablefoottext{b}{Exposure time.}
PI = Polarized Intensity; I = total Intensity; P = linear Polarization fraction}
\end{table*}

\subsection{Data Acquisition and Reduction}


Here we present archival scattered light observations of PDS 453. Specifically we analyze SPHERE polarized intensity and NICMOS total intensity and polarization fraction maps (see Table \ref{tab:observations} for observing details). All of these observations trace small grains in the upper surface layers of the disk.

\subsubsection{VLT/SPHERE Polarized Intensity Image}

PDS 453 was observed with the SPHERE instrument \citep{Beuzit2019} at ESO/VLT in the differential polarimetric imaging mode \citep[see][]{vanHolstein2020,deBoer2020} with the IRDIS camera \citep{Dohlen2008}. The full polarimetric imaging sequence was obtained in the broad band H-filter, as part of SPHERE's GTO programme 1100.C-0481(R). PDS 453 was observed for a total of 2304\,s on-source (38.4\,min), separated into 24 individual exposures of 96\,s. The atmospheric conditions were favorable, with a seeing in the range 0\farcs6-0\farcs7 for the complete sequence. SPHERE's apodized-pupil Lyot coronagraph \rev{mask} N\_ALC\_YJH\_S \rev{with an 0\farcs1 radius} was used to block the light from the central source 

The data reduction was performed with the IRDAP pipeline\footnote{\url{https://irdap.readthedocs.io}} \citep[see][for details on the observing procedure and data reduction process]{vanHolstein2020}. In short, the pipeline first performs the basic steps of data reduction (dark subtraction, flat fielding, bad-pixel correction and centering). Then, Stokes $Q$ and $U$ frames are calculated using the double difference method. The data are then corrected for instrumental polarization and cross-talk by applying a detailed instrument model (a Mueller matrix) that takes into account the complete optical train of the telescope and instrument. From the final $Q$  and $U$  images, a linear polarized intensity (PI) image is finally obtained \rev{by ${\sqrt{Q^2 + U^2}}$}. The total intensity images were not used because the bright central star leaves significant PSF residuals \rev{and the procedure did not allow for a clean retrieval of the total intensity, thus} we focus here on the polarized intensity alone.
This final image is corrected for true north \citep{Maire2016}. The SPHERE PI image of PDS 453 is shown in the left panel of Fig. \ref{fig:OBSERVATIONS}. 

\subsubsection{HST/NICMOS Total Intensity Image and Polarization Fraction Map}

PDS 453 was observed using HST NICMOS coronagraphy as part of program 11155 (PI: Perrin). 
Following the recommended observational strategy for NICMOS coronagraphy, data was obtained in two roll angles to allow \rev{minimize} of instrumental artifacts. 
At each roll angle, the target was centered behind the NIC2 coronagraph occulter, images were taken using the three 2.0 $\mu$m linear polarizers (POL0L, POL120L, POL240L; spectral range 1.89-2.1 microns) and subsequently with the F110W filter (spectral range 0.8-1.4 microns). 

Data reduction followed methods previously described in \citet{2005ApJ...629L.117S} and \citet{2009ApJ...707L.132P}, including basic pipeline reductions, detection and correction for outlier bad pixels, thermal background subtractions, and PSF subtraction. To remove stellar PSF residuals, classical PSF subtraction was performed using a small library of PSF reference stars observed in programs 10847 and 11155. For the PDS 453 data, the best subtractions were obtained using PSF references GJ 273 and HD 21447 for the F110W and POL*L data, respectively. Following PSF subtraction, remaining artifacts such as diffraction spikes were masked out, images were rectified for geometric distortion, rotated to a common orientation 
For the \rev{NICMOS polarization data set}, the POLARIZE software \rev{transform the three POL*L images into $I$, $Q$ and $U$  images} \citep{2000AAS...197.1210H} 
The polarized images were smoothed by a one-resolution-element Gaussian to reduce noise. The linear polarization fraction was computed in the usual fashion as ${\sqrt{Q^2 + U^2}}/{I}$ from the Stokes images. \rev{The two images at both roll angles were average together once rotated so that North is at the top of the image.} 

\begin{center}
    \begin{figure*}[h!]
        \centering
        \includegraphics[width=1\textwidth]{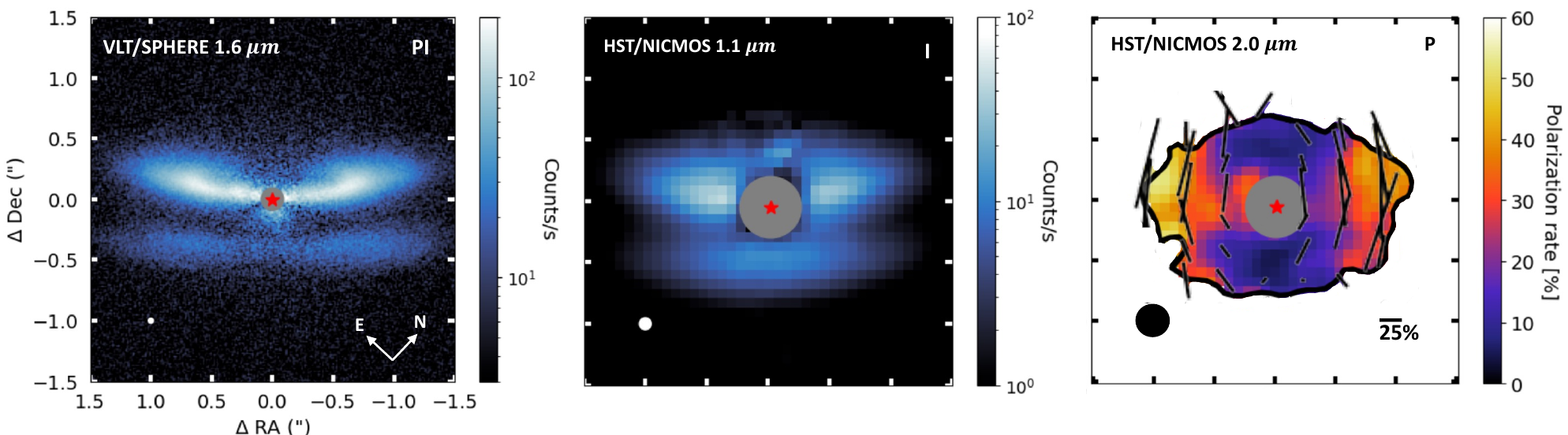}
        \caption{High-resolution observations of PDS 453. The size of the corresponding PSF is shown at the bottom left of each panel and the gray circles represent the size of coronagraphics masks. \final{The red star symbol corresponds to the position of the bright central point source.} \textbf{\textit{Left panel:}} VLT/SPHERE 1.6\,$\mu$m polarized intensity image shown with a logarithmic stretch. \textbf{\textit{Middle panel:}} HST/NICMOS scattered light F110W total intensity image shown using a logarithmic stretch. \textbf{\textit{Right panel:}} HST/NICMOS scattered light 2\,$\mu$m polarization fraction map along with polarization vectors. The polarization fraction is only computed in the area where the total intensity is detected at the 1\,$\sigma$ level. All images have the same scale, $3\times3$ arcsec.}
        \label{fig:OBSERVATIONS}
    \end{figure*}
\end{center}

The resulting 1.1\,$\mu$m total intensity image and 2\,$\mu$m polarization fraction map are shown in Fig. \ref{fig:OBSERVATIONS} (middle and right panels). \rev{The total intensity and polarized intensity images at 2\,$\mu$m are shown in Fig. \ref{fig:hst_2_micron}}. At the longer wavelength, the total intensity image is too contaminated by PSF subtraction residuals and we choose to only analyze the polarization fraction map. Despite its worse angular resolution, the polarization fraction map provides unique insight on dust properties when compared to the SPHERE PI image.

\subsection{Observational Results}

The SPHERE and NICMOS images presented here confirm the presence of a highly inclined disk around PDS 453 \rev{identified by \citet{2010AAS...21542812P}.} The outer radius of the disk is $\approx$1\farcs2 and a well resolved dark lane bisects the two disk surfaces. The curved ring-like nature of the upper surface indicates that the disk is inclined at $\sim$80\deg. Furthermore, this suggests a sharp transition in the disk density profile at a radius of \rev{$\approx$0\farcs5 ($\sim$70 au) from the star, similar to 2MASS J16083070-382826 \citep{Villenave_2019}}.

\begin{center}
    \begin{figure*}[h!]
        \centering
        \sidecaption
\includegraphics[width=0.55\textwidth]{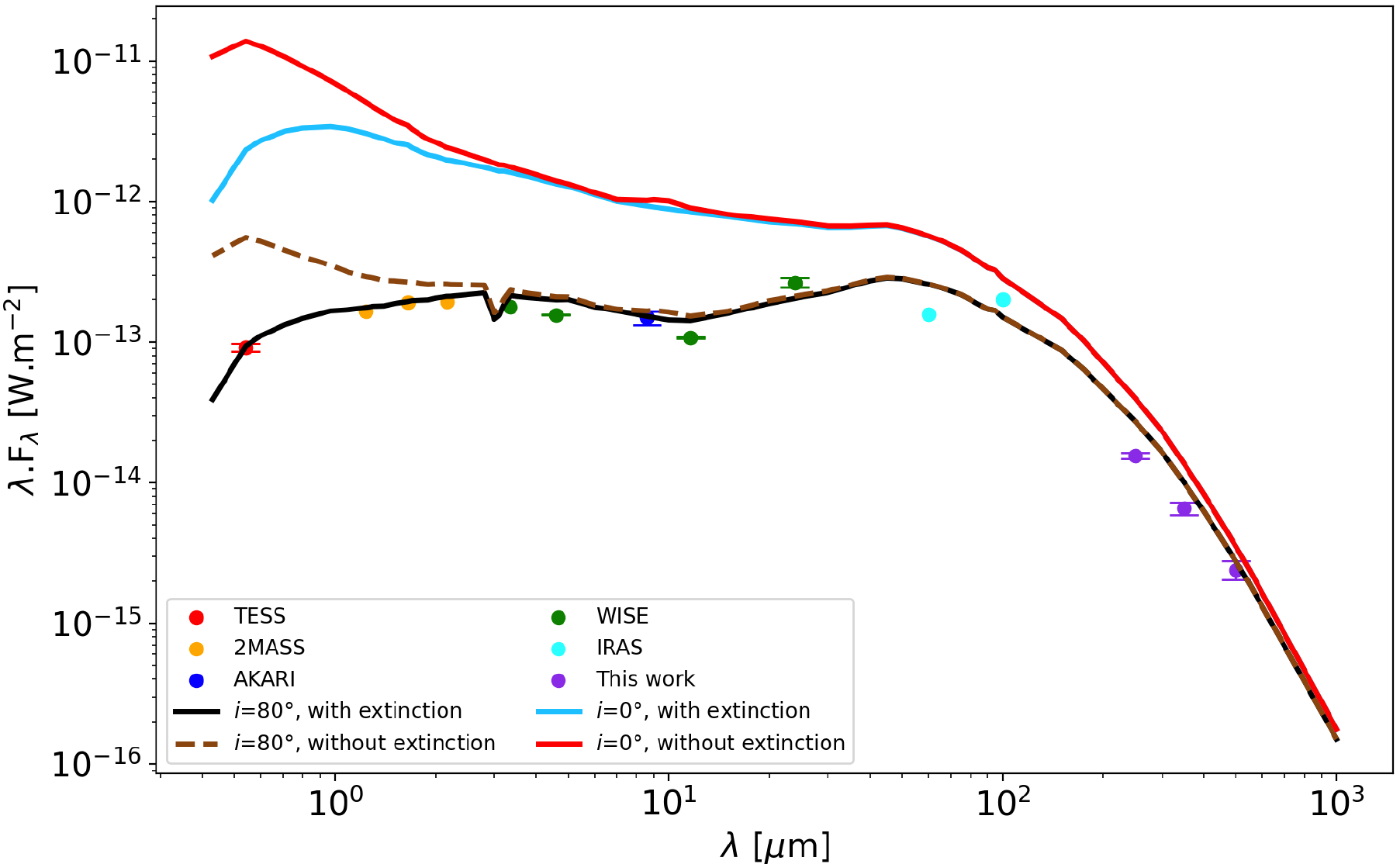}
        \caption{SEDs of the best model of PDS 453 with and without foreground extinction (solid black line and dashed brown line with $A_V=1.9$\,mag, respectively). The SEDs of the same model viewed pole-on are shown with and without the same visual extinction (blue solid line and red dashed line, respectively). Upper limits are indicated with downward-pointing arrows. See Section\,\ref{sec:results} for models description.}
        \label{fig:best_SED}
    \end{figure*}
\end{center}

A bright central point source is observed along the top surface, a rare configuration among such highly inclined disks. \rev{This bright central point source is not visible in our observations because it was blocked by the coronagraph and subtracted as part of the data reduction process.} To determine the nature of this component, we compiled the SED of the system. To this end, we complement existing optical through mid-infrared photometry from all-sky surveys available in \textit{VizieR}\footnote{\url{http://vizier.cds.unistra.fr/vizier/sed/doc/}} with new far-infrared Herschel photometry. PDS 453 is included in one of the wide-field SPIRE maps taken as part of the Herschel Gould Belt Survey Guaranteed Time Key Programme \citep{andre2010}. We retrieved the level 2.5 SPIRE (250, 350 and 500 $\mu$m) data products from the Herschel archive. 
PDS 453 is clearly detected in all three filters. Following the SPIRE Handbook, we estimated photometry for the system using Gaussian fitting and applying band-appropriate aperture corrections. This yielded flux densities of 1.30$\pm$0.06, 0.77$\pm$0.08, and 0.40$\pm$0.06\,Jy at 250, 350 and 500\,$\mu$m, respectively. The resulting SED is shown in Figure \ref{fig:best_SED}. The optical photometry shows that the system is underluminous in the visible and near-infrared compared to other F-type stars in Sco-Cen \citep{Pecaut_2012}. This indicates that the central point source is a forward scattering glint through the upper layer of the disk rather than a direct view of the star. 

Moreover, the SED of the system is typical of so-called "flat spectrum" sources \citep{2008A&A94}. While these objects are generally understood as an intermediate evolutionary stage between embedded and revealed young stellar objects, edge-on disks around T\,Tauri stars can yield similarly-shaped SEDs depending on their exact orientation, \citep[e.g.,][]{Glauser2008}. The images of PDS 453 do not indicate the presence of a remnant envelope and thus we interpret it as a similar object. In other words, the system is viewed at a peculiar angle, at a grazing incidence through the upper layers of the disk, which enables us to directly trace its surface.

The 2\,$\mu$m polarization fraction measured with NICMOS increases monotonously from $\sim$20$\%$ at the edge of the coronagraphic mask to $\sim$50$\%$ near the edge of the disk. We estimated these fractions by averaging the total and polarized intensity signals between two horizontal lines that bracket the spine of the top nebula. 
The orientation of the polarization vectors is orthoradial in all regions, consistent with scattering of \rev{stellar photons off dust grains} (see right panel of Fig. \ref{fig:OBSERVATIONS}).

\section{Modeling Setup}
\label{sec:modeling}

In this section, we aim to build a model that reproduces the various observational elements established in the previous section, namely the system SED, the 1.6\,$\mu$m polarized intensity image and the 2\,$\mu$m polarization fraction map. 
While the model was not optimized for the 1.1\,$\mu$m NICMOS total intensity image, it provides an additional constraint on the dust properties and is included for discussion. 

We used the MCFOST radiative transfer code \citep{Pinte2006} which produces synthetic SEDs, scattered light images and polarization maps and assumed that the disk is axisymmetric. The model exploration started from a setup designed by \citet{2010AAS...21542812P} that we modified to improve the global adjustment, in particular match the higher resolution of the SPHERE image. We caution that we did not attempt to obtain the best quantitative match to the data, leaving to a further study such detailed model fitting. Nonetheless, as shown below, we achieve a satisfactory match to all key aspects of the \rev{selected} observations. 

We fixed the disk inclination to 80\degr\ based on the morphology of the SPHERE image. We \final{used a photospheric spectrum to represent the star \citep{Baraffe_1998} and} set the effective temperature of the star to 6810\,K, appropriate for an F2-type star \citep{Pecaut_2013}, and the stellar luminosity to \final{8.69\,$L_\odot$, typical} of early-F stars in the nearby Upper Centaurus Lupus star forming region \citep{Pecaut_2012}.  

\begin{center}
    \begin{figure}[h!]
        \centering
        \includegraphics[width=0.5\textwidth]{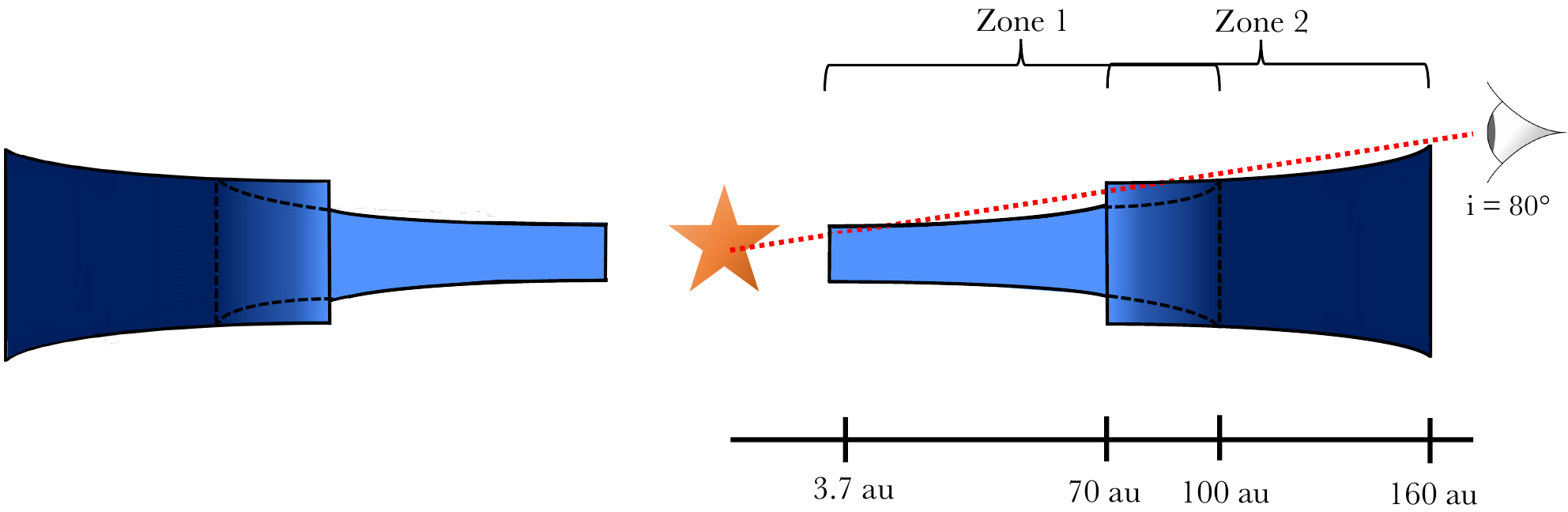}
       \caption{Sketch of the PDS 453 model viewed at 80°.}
        \label{fig:sketch}
    \end{figure}
\end{center}

We initially considered a simple model, with a single surface density profile extending smoothly to the disk outer radius. However, this model \rev{produced a smooth image without the sharp ring seen in the SPHERE observation} (see Fig. \ref{fig:model_1_zone}). We therefore designed a model with two partially-overlapping regions, with the outer one characterized by a higher scale height and a higher degree of flaring. This combination was necessary to reproduce the scattered light polarized intensity distribution. We note that the use of two distinct regions is often necessary to reproduce observations of protoplanetary disks \citep[e.g.,][]{2019PASP..131f4301W, 2023A&A...672A..30K}. Fig. \ref{fig:sketch} presents a sketch of the geometry we adopt for the PDS 453 disk.

We adopted a power law distribution, $\Sigma(R) \propto R^p$, with sharp edges for the surface density profile in the inner region. Conversely, we needed to use a tapered-edge profile, $\Sigma(R) \propto R^p\exp( - (\frac{R}{R_c})^2)$ where $R_c$ is a so-called critical radius, in the outer region. We assume a Gaussian vertical density profile, appropriate \rev{for} a vertically isothermal disk in hydrostatic equilibrium. The disk scale height is parametrized as $H(R) = H_{0}\left(\frac{R}{R_{0}}\right)^{\beta}$, where $\beta$ is the flaring index and $R_0$ is an arbitrary reference radius which we set to 135 au. Each region has its inner and outer radius, $R_{in}$ and $R_{out}$, and dust mass, $M_{d}$. 


We assume that all grains are spherical and homogeneous, allowing us to use Mie theory. \rev{While it is numerically convenient, Mie theory is known to be imperfect in reproducing scattered light images \citep[e.g.,][]{2012A&A...537A..75M}, as grains likely have more complex structures \citep{Pinte2008,Tazaki2023}. It is nonetheless widely used in the field \citep[e.g.,][]{Birnstiel2018} and we adopt it as it does not qualitatively alter our conclusions. The grains} are composed of a mixture of two solid species: "dirty" water ice mostly crystalline as described in \citet{1998A&A...331..291L}, and a  mixture of graphite, silicate and amorphous carbon developed by \citet{1989ApJ...341..808M} to match interstellar dust properties. We apply Bruggeman rule effective medium theory to compute the effective refractive index, including an additional level of porosity. We explored different proportions for the two species, with an ice volume fraction ranging from 5 to 50\% in the outer disk region. In the inner region, we consistently assumed a lower ice proportion (by 10\% in an absolute sense, or by half, whichever is smaller) than in the outer one. Finally, close to the star the disk is warm enough to sublimate water ice. In a preliminary model, we found that the midplane temperature is higher than 90\,K inside of 3.7\,au, and we therefore removed all water ice inside of that radius for all models. 
\rev{To determine the depth the water ice band, we normalized the spectrum by a second or third-order polynomial fit to the 1.65–2.3\,$\mu$m and 4.0-4.6\,$\mu$m continuum regions and obtain the relative depth of the band from its minimum.}

We assume a power law grain size distribution $n(a)da \propto a^{-q}da$ with $q=3.5$ extending from $a_{min}=0.03\,\mu$m to $a_{max}=3000\,\mu$m, where the maximum grain size was set so as to match the long wavelength end of the SED. Dust grains are assumed to be in radiative equilibrium and local thermodynamic equilibrium, appropriate for an optically thick disk. 
            
\begin{table*}[h!]
    \centering
    \caption{Parameters of PDS 453 model. LTE: Local Thermal Equilibrium. EMT: Effective Medium Theory.}
    \renewcommand{\arraystretch}{1.2}
    \begin{tabular}{cccccc|ccc}
        \toprule
        \toprule
        \rowcolor{gray!20} & \multicolumn{4}{c}{\centering System parameters} && \multicolumn{3}{c}{\centering Star parameters} \\
        \midrule
        Phase function & Heating & Mixing & Porosity & Inclination & Distance & Temperature & Luminosity & Radius\\
        of diffusion & method & rule & ($\%$) & $i$ (°) & $(pc)$ & $(K)$ & ($L_{\odot}$) & ($R_{\odot}$)\\
        \midrule
        Mie & LTE & EMT & 50 & 80 & 130 & 6810 & \final{8.69} & \final{2.12}\\
    \end{tabular}
    
    
    \begin{tabular}{cccccccccc}
        \toprule
      \rowcolor{gray!20} \multicolumn{9}{c}{\centering Disk parameters} \\
        \midrule
        Zone & Type & R$_{in}$ - R$_{out}$ & Ratio & Dust mass & Scale height & Flaring & Surface density & Grain size \\
        & & (au) & ($\text{dust}:\text{ice}$) & ($M_{\odot}$) & $H_{0}$ (au) & exponent $\beta$ & exponent $p$ & ($\mu$m) \\
        \midrule 
        Inner  & Sharp edges & 0.2 - 100 & 0.9:0.1 & \num{8.83e-6} & 8.50 & 1.13 & -1.00 & 0.03 - 3000 \\
        Outer & Tapered edge & 70 - 160 & 0.8:0.2 & \num{9.2e-6} & 11.23 & 1.30 & -0.25 & 0.03 - 3000 \\
        \bottomrule
        \bottomrule
    \end{tabular}
    \label{tab:parameters}
\end{table*}

As we compared the resulting model images and SED to the observed ones, we varied the mass, scale height, flaring index, surface density index and radial extent of both regions, as well as the water ice fraction \rev{through a manual exploration.} All the model parameters (fixed and varied) are summarized in Table \ref{tab:parameters}.

\section{Results}
\label{sec:results}


\subsection{Disk Morphology}

\begin{center}
    \begin{figure*}[h!]
        \centering
        \includegraphics[width=1\textwidth]{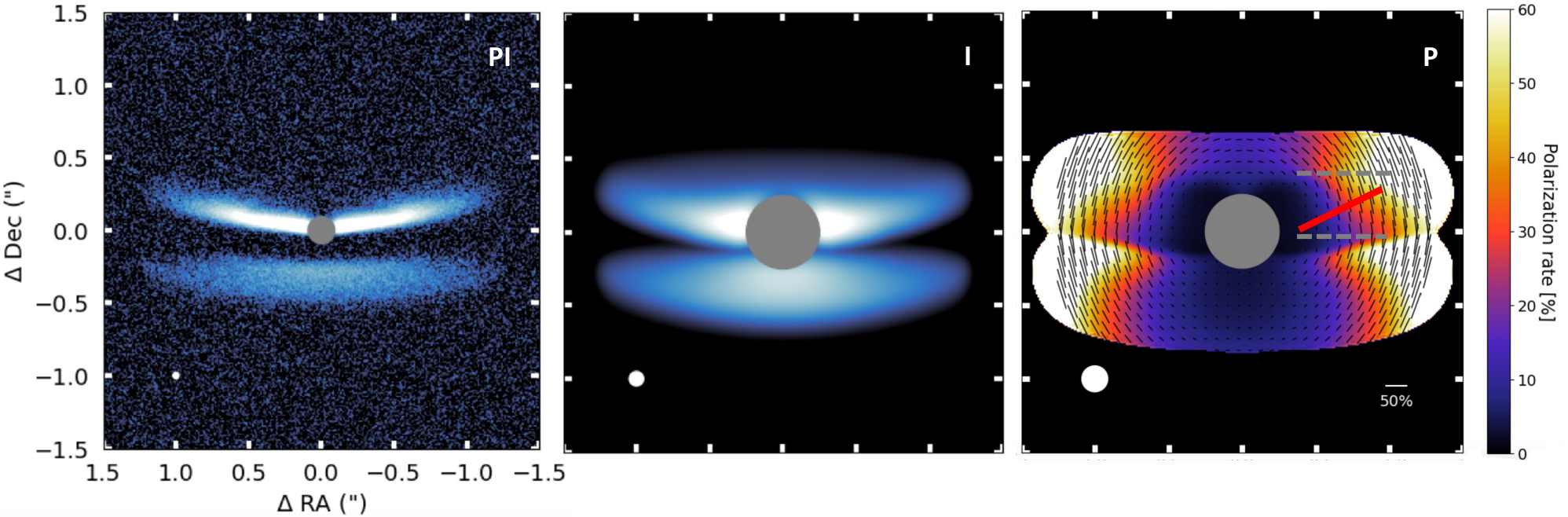}
       \caption{Synthetic observations of our best PDS 453 model \rev{including two zones}, with similar scales and stretches as in Fig.\,\ref{fig:OBSERVATIONS}. All images have been convolved by the appropriate PSF after subtracting the central point source to mimic the effect of the coronagraph.
       \textbf{\textit{Left panel:}} Scattered light 1.6\,$\mu$m polarized intensity image shown a the SPHERE resolution. \textbf{\textit{Middle panel:}} Scattered light F110W total intensity image at the NICMOS resolution. \textbf{\textit{Right panel:}} Scattered light 2\,$\mu$m polarized fraction map with polarization vectors superimposed. The red solid line approximates the location of the spine of the top nebula. The two gray lines represent the vertical range over which the total and polarized intensities are measured to compute  the polarized fraction shown in Fig. \ref{fig:polar_rate_evolution}.}
        \label{fig:MODELS}
    \end{figure*}
\end{center}

The synthetic images and polarization map of our final model are presented in Fig. \ref{fig:MODELS}. The model satisfyingly matches all observations in \rev{terms of morphology (radius, aspect ratio, presence of a bright ring), width of dark lane, brightness ratio between the 2 nebulae, size and maximum polarization}. The outer disk radius is 160\,au, as indicated by the extent of the SPHERE scattered light image. The curved upper surface (above the darklane in Fig. 1) and the flatter nature of the lower surface (below the darklane), are well reproduced. As discussed \rev{in Sect. \ref{sec:ObsRes}}, the turnaround points towards the back side on the top surface, suggest a ring-like structure, akin to a transition disk. However, the SED of the system shows infrared excess extending from 2\,$\mu$m to the submillimeter, so that a fully cleared inner disk is ruled out. We note that the temperature of water ice sublimation is reached at $\sim$60 au at the surface of the disk in our model, suggesting that this transition could be due to the location of snow line in the system. The lack of polarized intensity on the back side of the ring is well matched, so long as porosity is included in our model.

The main shortcomings of our model are related to the lower nebula. In the SPHERE observation, the polarized intensity is fainter in the center than at the edges of the disk whereas in the NICMOS 1.1\,$\mu$m image, its radial extent is shorter than the upper one. Neither features are well reproduced (see Fig.~\ref{fig:MODELS}), suggesting that the scattering phase function and/or polarizability curves of the dust model we adopted is imperfect\rev{, possibly due to our choice of using Mie theory}. \rev{Moreover, the HST and SPHERE observations being separated by 10 years, another possible explanation is the change in illumination, for instance as the result of a misaligned inner disk \citep[e.g.,][]{benisty2022optical}.}

Given our grazing angle perspective to the system, the disk surface is traced primarily by single scattered photons. Still, the disk blocks our direct view of the star, such that the bright central point source we see is substantially fainter than the star itself. In our model, the central point source when the system is observed at 80\degr\ inclination has $\sim$3$\%$ of the brightness of the star at 1.1 $\mu$m (as measured in a pole-on view of the same model). 

Very few protoplanetary disks match this peculiar geometry. Most prominently, MY Lup, with an inclination of $i = 73$\degr, also presents a bright inner ring feature, a dark lane, and a bright central point source \citep{Avenhaus2018}. In that system, \citet{Jennings_2022} demonstrated that the ring feature coincides with a trough in the sub-mm continuum emission. Unfortunately, no resolved submillimeter observations of PDS 453 exist to date. A few additional systems in other star-forming regions present similar morphologies: DoAr 25, V1012 Ori, PDS 111, V409 Tau and RY Tau \citep{2020A&A...633A..82G, 2024A&A...685A..54V, 2024arXiv240604160D, 2024A&A...685A..53G}. These could represent a valuable coherent sample for future systematic studies.

\subsection{Spectral Energy Distribution}
\label{sec:results:sed}

The SED of the model \rev{(Fig. \ref{fig:best_SED})} matches well the observed one if we include a foreground extinction of $A_V=1.9$\,mag (assuming an extinction law characterized by $R_v=3.1$). The extinction estimate we obtain is consistent with the low galactic \rev{latitude} of PDS 453. \rev{Indeed, \citet{Esplin_2020} estimate an extinction of $A_V=2-3$\,mag for the stellar population of the nearby Upper-Scorpius star forming region}. This is significantly higher than the extinction of $A_v=0.1$\,mag estimated by \citet{Sartori_2010}. This difference is due to the fact that scattering off the circumstellar medium does not preserve the colors of the central source and therefore biases the extinction estimates. 

The spectral index between 350 and 1000\,$\mu$m of the model is estimated to be $\alpha = -2.1$, using $F_\nu \propto \nu^\alpha$. This suggests that the disk is optically thick in this regime. Longer wavelengths observations \rev{in the millimetric regime} are required to estimate this index empirically and confirm this conclusion.

The model presented here produces a relatively shallow water ice absorption band around 3.1 $\mu$m, in qualitative agreement with the observations of \citet{Terada_2017}. We note that the foreground extinction is too low to be expected to produce water ice absorption where \rev{$A_V$ must be $\gtrsim$ 3 mag according to} \citet{2015ARA&A..53..541B} and that there is no evidence of a remnant envelope associated with the system. We conclude that the water ice \rev{band} originates from the disk itself.

\subsection{Water Ice Absorption \rev{Band}}
\label{sec:results:water}

\rev{Irrespective of whether absorption or scattering dominates, the peculiar geometry of PDS 453 is such that we can conclude that water ice is present in disk layers located much above the midplane. This is despite the fact that in this layer water ice could easily sublimate over a large fraction of the disk radial extent.} Water ice absorption is usually attributed to an effect of transmission of starlight through foreground material. However, in the case of highly inclined disks where scattering dominates, the dust albedo spectrum becomes relevant and can also contribute significantly to the 3 $\mu$m \rev{water} ice \rev{band}. 

Our model produces a clear water ice \rev{band (Fig. \ref{fig:best_SED})}, although it does not perfectly match the observed one. Discrepancies in the detailed \rev{shape and central wavelength} of the profile could arise from the assumed type of water ice. Here we considered amorphous ice while \citet{Terada_2017} use highly crystalline ice. As shown in Fig\,14 of \citet{2021ApJ...921..173T}, there are significant differences in the refractive indices of these two types of ices. Moreover, \citet{Terada_2017} measured a relative depth of the minimum of the \rev{water} band of $\sim$24$\%$, taking into account the bias introduced by the limited wavelengths range from their continuum estimation. 
In our model, the \rev{water} ice band is deeper, with a depth of $\sim$33$\%$ relative to the continuum. A slightly lower inclination or ice content could help improve this aspect, see \S~\ref{sec:WIAB}, but it would come at the cost of poorer matches to the other observable quantities.

Finally, we note that the ice volume fraction used in our PDS 453 model \rev{10-20$\%$} is similar to what \rev{is typically suggested from water ice observations for disks across a wide range of inclinations \citep{2005ApJ...622..463P, 2021ApJ...921..173T, 2023A&A...677A..18S, Sturm_2024}}. 
\rev{This value is significantly lower than the one often used in dust opacity modeling in protoplanetary disks \citep{Pollack_1994, Dalessio_2001, Birnstiel2018}, 
where a water ice volume fraction of about 36-60$\%$ is assumed based on the solar abundances or cometary observations. The relatively low ice abundances in the disk surface might point to disruption of water ice via non-thermal processes \citep{Hollenbach_2009, Oka_2011}, although a detail physical-chemical modeling \citep[e.g.,][]{Ballering_2021} is needed to clarify the origin.}

\subsection{Polarization Fraction}

The 2\,$\mu$m polarization fraction increases with distance from the star in the model (Fig. \ref{fig:MODELS}, right panel), in agreement with the measured polarization fraction from the NICMOS data (Fig. \ref{fig:OBSERVATIONS}, right panel). This effect is due to a scattering angle effect: close to the star, the light is perceived as forward-scattering, while near the outer edge of the disk, the light is being scattered at 90\degr\ scattering angle. The latter generally produces higher polarization fractions. The pattern of polarization vectors in the model is centro-symmetric, consistent with the observations and with previous studies of nearly edge-on young stellar objects  \citep{2010A&A...518A..63M}. 

The general behavior is \rev{satisfactorily} reproduced by our model \rev{with a continuous rise from the inside to the outside of the disk as well as the correct maximum polarization at the edge of the disk $\approx40\%$. However, the model under-predicts the polarization fraction at small projected separations (see Fig. \ref{fig:polar_rate_evolution})}.
Close to the star, this difference may be caused by imperfectly subtracted starlight leakage at the edge of the coronagraph and/or by the intrinsically biased nature of the polarization fraction. \rev{The difference could arise from a density profile with too sharp an edge compared to observations. Moreover, 
no noise was added in the HST images model.} 
\rev{The difference in radius between the observations and our model arises from the fact that $R_{out}$ of the model is based on the dimension of the disk as seen in the SPHERE image. However, the polarization map coming from the 2 $\mu$m NICMOS image, we defined the limited radius of the observations based on the detection limit of the 2 $\mu$m total intensity scattered light image presented in Fig. \ref{fig:hst_2_micron}.}
\section{Discussion}
\label{sec:discussion}

The model we have constructed reproduces the observations of PDS 453 despite small shortcomings and unsolved degeneracies inherent to such an exercise. In the absence of \rev{higher} resolution observations, in particular with ALMA (a good tracer of the total \rev{mass of mm dust}), we consider that it is premature to try and refine the model to fit the data with more accuracy. Nonetheless, we consider that the two-zone nature of our model, as well as the presence of water ice in the upper disk layers that absorbs and/or scatters light from the central star, are \rev{directly} established \rev{from observations}. Beyond the properties of this disk, we take advantage of our model to explore more generally how the water ice band and the polarization fraction are dependent on physical parameters, such as viewing angle, disk vertical structure and dust properties.

\subsection{Water Ice Absorption \rev{Band}}
\label{sec:WIAB}
\subsubsection{Sensitivity to Inclination}

\begin{center}
    \begin{figure}[h!]
        \centering
        \includegraphics[width=0.45\textwidth]{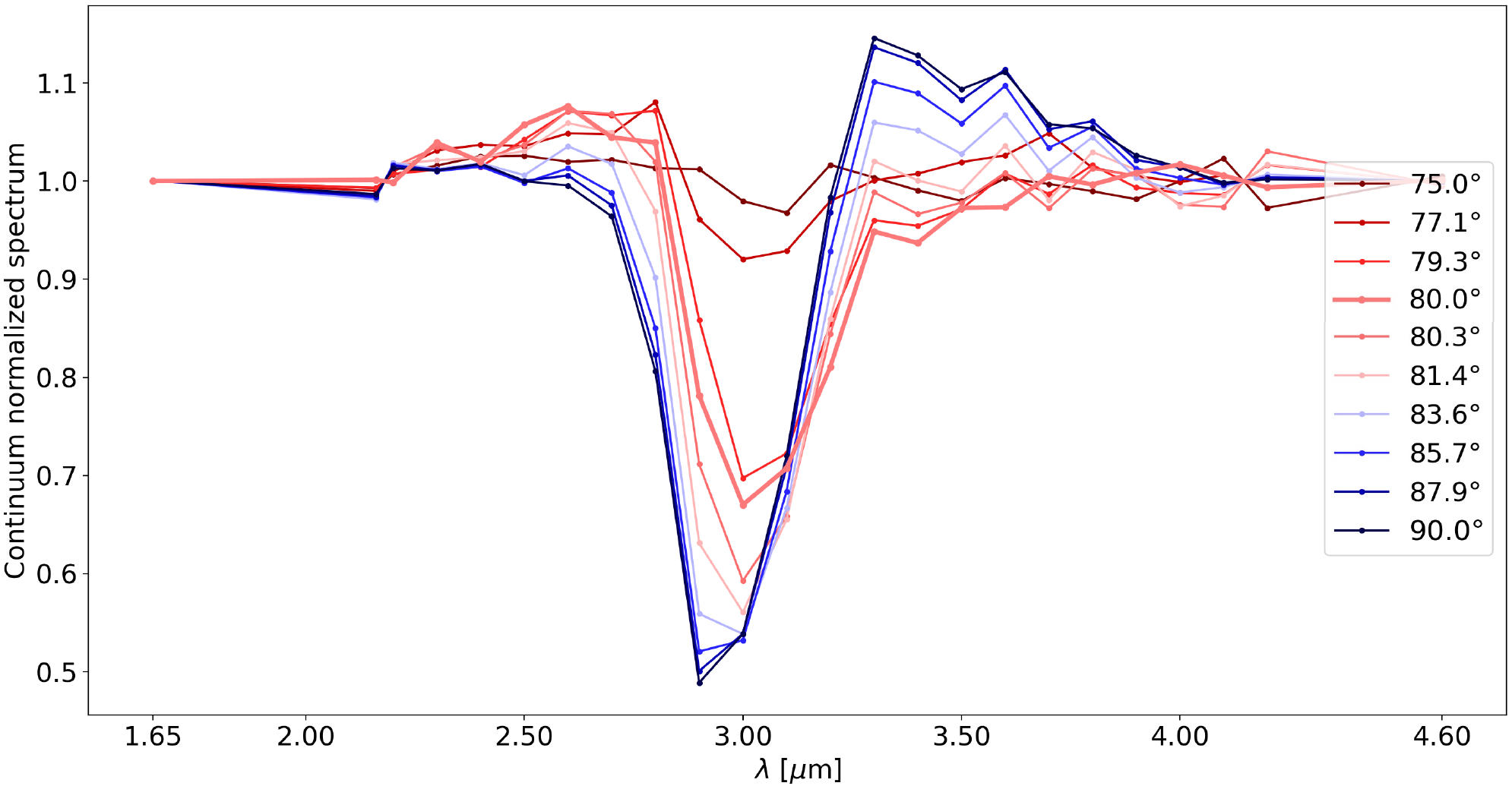}
        \caption{\final{Continuum normalized} synthetic model spectra centered around the water ice band as a function of inclination. We only present every other inclination for clarity. 
        \rev{The spectrum at 80° represents our best model of PDS 453.}} 
        \label{fig:inclinations}
    \end{figure}
\end{center}

We now explore the influence of the system inclination on the depth and spectral shape of the water ice \rev{band}. To this end, we started from our best PDS 453 model \final{including star and disk} and computed the emergent spectrum in the range 1.6 -- 4.6\,$\mu$m, while varying the inclination from 75 to 90\degr, where $i=80$\degr\ corresponds to the PDS 453 geometry (see Fig. \ref{fig:inclinations}). 
For \final{$i<75$\degr}, the \rev{stellar} photosphere of the central star is directly visible and no \rev{water} ice \rev{band} is detected. Above this limit, in addition to an increase in the depth of the \rev{water ice} band, the spectral profile of the band varies with inclination. The minimum of the band shifts from $\sim$3.1 $\mu$m $\sim$to 2.9 $\mu$m as inclination increases. Furthermore, the \rev{water ice} band is asymmetrical with a wider "wing" that switches from being on the blue to the red side of band. The \rev{water ice} band is almost symmetrical at $\sim$82\degr. Finally, there is a small "bump" over the continuum which also switches from the blue to the red side as inclination increases.

\begin{center}
    \begin{figure*}[h!]
        \centering
        \includegraphics[width=0.9\textwidth]{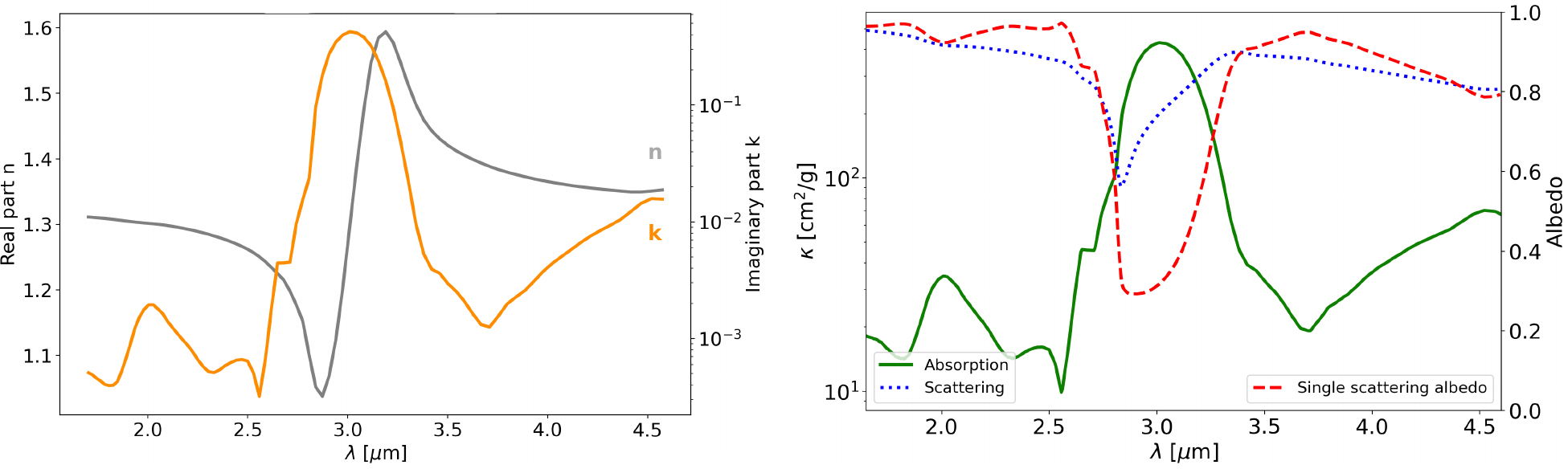}
        \caption{\textbf{\textit{Left panel :}} Refractive indices of water ice from \citet{1998A&A...331..291L}. The real part is displayed on a linear scale and the imaginary part on a logarithmic scale. 
        \textbf{\textit{Right panel:}} Opacities on a logarithmic scale and albedo of water ice on a linear scale.}
        \label{fig:refractive_opacities}
    \end{figure*}
\end{center}

The refractive index of water ice provides an explanation to this behavior with inclination. Fig. \ref{fig:refractive_opacities} shows the refractive index and opacity of water ice. Broadly speaking, the real and imaginary parts are linked to scattering and absorption, respectively. At intermediate inclinations (78--82\degr), where the optical depth through the disk is not negligible but moderate, absorption dominates over scattering and we essentially find a transmittance spectrum. Given the absorption opacity curve, the \rev{water} ice band is centered at $\gtrsim 3.0\,\mu$m, as observed in our model. At even higher inclinations, the optical depth becomes so high that no photons are directly transmitted to the observer. The spectrum is then dominated by scattering. Since the minimum albedo occurs at $\approx2.9\,\mu$m, this explains the shift to shorter wavelengths in our model. This behavior is in agreement with \citet{2015ARA&A..53..541B} who showed that detection of the \rev{water ice} band at 2.9\,$\mu$m is indicative of scattering. In the case of PDS 453, the \rev{water} ice band is located around 3.1\,$\mu$m \citep{Terada_2017} which is therefore primarily linked to the absorption by water ice, in agreement with the line-of-sight to the photosphere being at grazing incidence on the disk surface, i.e., moderate optical depth at 3.1 $\mu$m. 





Asymmetry in the \rev{water ice} band wings may be explained in a similar manner.
The blue wing at low inclinations is influenced by the wavelength dependency of the optical constants of the materials \citep{2023NatAs...7..431M}, whereas the asymmetry of the red wing is due to a strong scattering effect by large grains \citep{1983A&A...117..164L, 2015ARA&A..53..541B}. The fact that large grains produce such an effect means that a distortion in ice bands may be an indicator of grain growth, with sizes well above the ISM \citep{2024NatAs.tmp....6D}. 


\subsubsection{Sensitivity to Water Ice Abundance}

\begin{center}
    \begin{figure}[h!]
        \centering
        \includegraphics[width=0.45\textwidth]{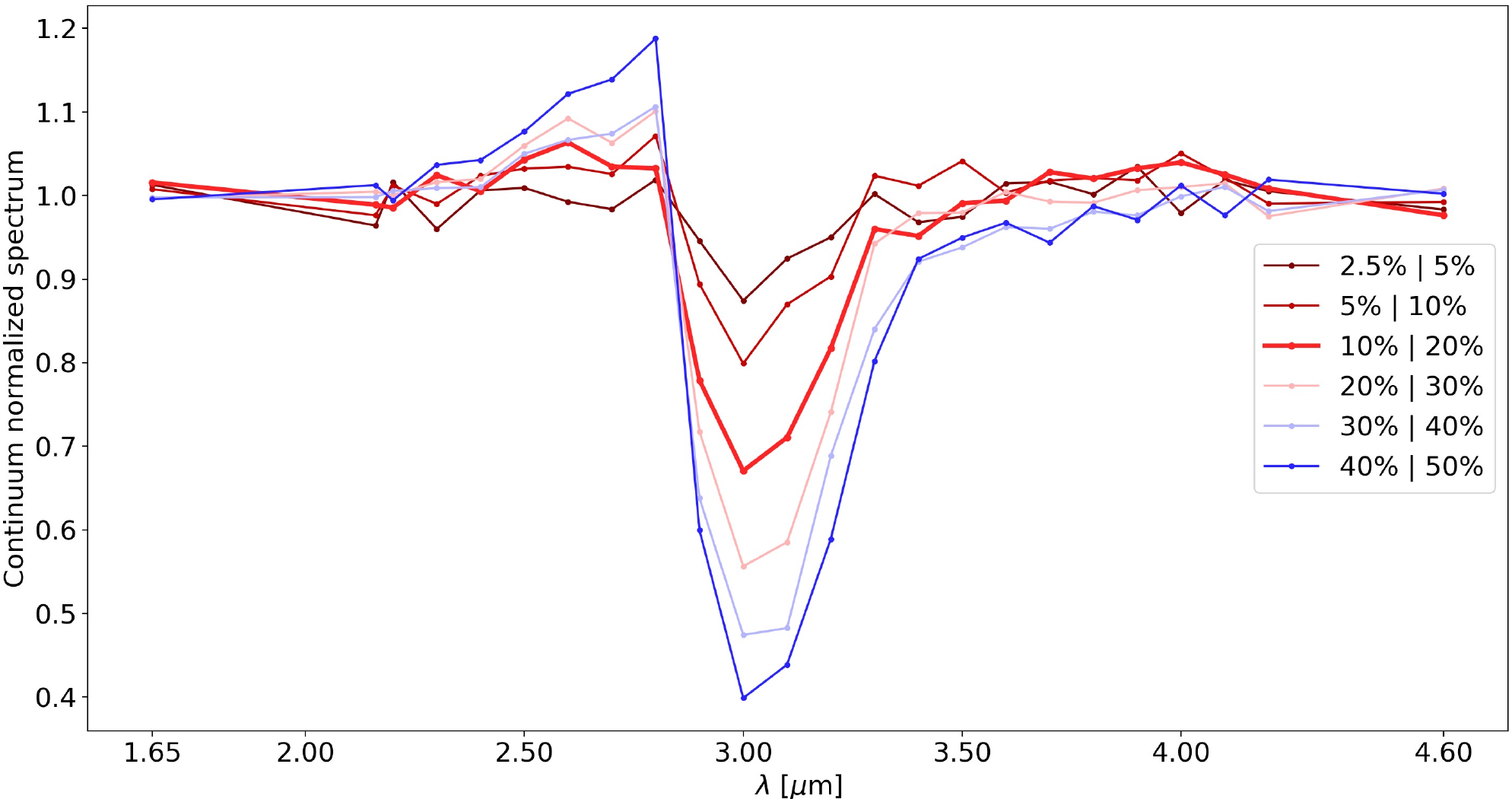}
        \caption{\final{Continuum normalized} synthetic model spectra centered around the water ice band as a function of volume fraction of ice in the disk. \rev{The spectrum with 10:20$\%$ of water ice represents our best model of PDS 453.}} 
        \label{fig:ice_quantity}
    \end{figure}
\end{center}

The amount of ice also has an influence on the depth of the water ice band. We varied the amount of water ice in the model and the results are shown in Fig. \ref{fig:ice_quantity}. Other than the water ice proportion in both regions, all other model parameters remain unchanged. The water ice band is detectable with the lowest volume fraction of ice we considered (2.5\% and 5\% in the inner and outer regions, respectively). In contrast to the variation of inclination, the wavelength of the \rev{water ice} band minimum does not depend on the ice content and remains at $\approx3.0\,\mu$m. The wings of the \rev{water ice} band are symmetrical for low ice abundances and become asymmetrical with a more pronounced blue wing for higher ice abundances. 


\subsubsection{Implication for the Interpretation of the Depth of the Water Ice Band}

\begin{center}
    \begin{figure}[h!]
        \centering
        \includegraphics[width=0.45\textwidth]{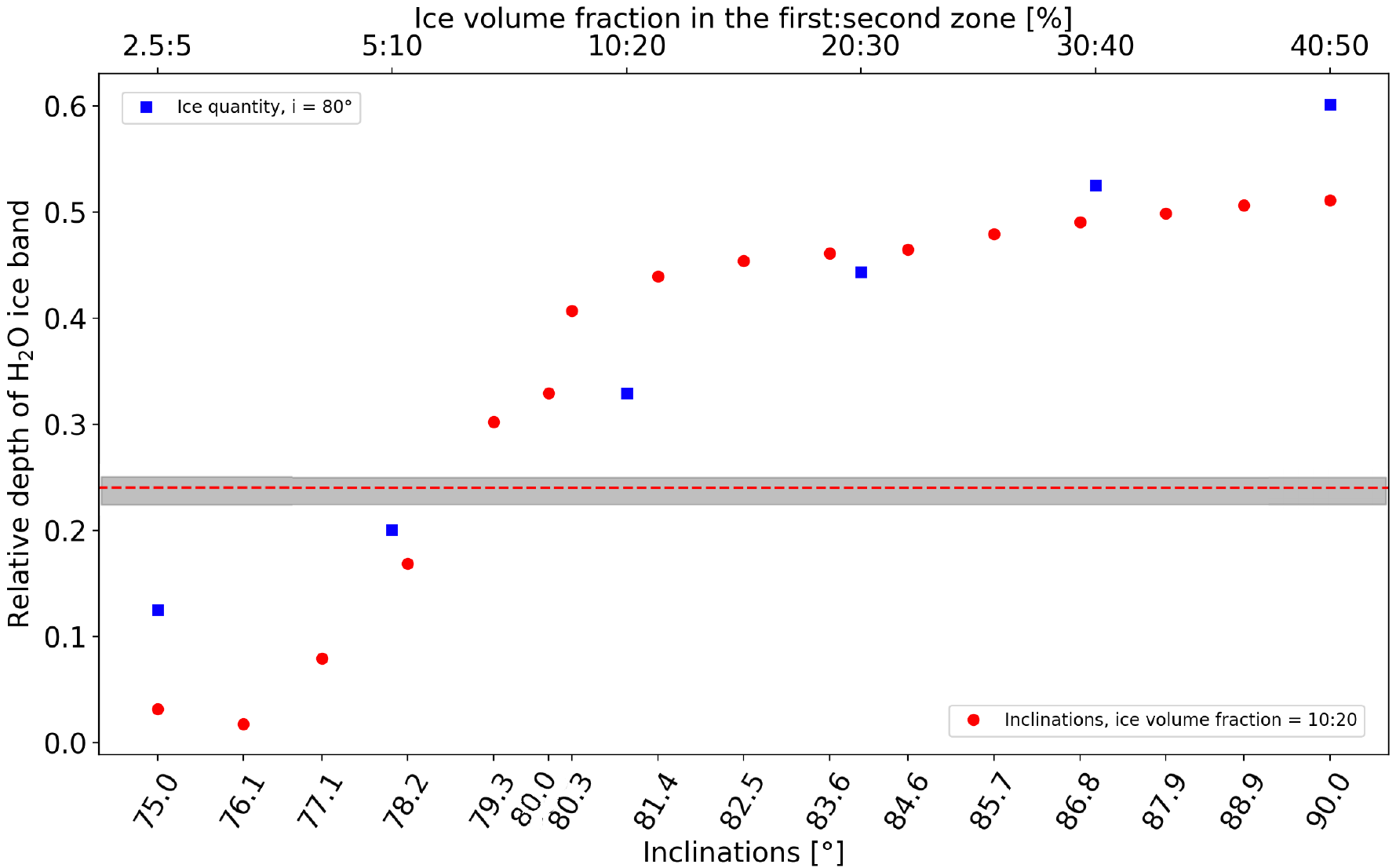}
        \caption{Variations of the relative depth of the water ice band as a function of the system inclination and the ice quantity. All model parameters remain unchanged excepted inclination for the \rev{red dots} and ice quantities for the \rev{blue dots}. The red horizontal \rev{dashed} line indicates the measured band depth \citep{Terada2017}.} 
        \label{fig:ICE_SATURATION}
    \end{figure}
\end{center}

Fig. \ref{fig:ICE_SATURATION} summarizes our findings about the dependency of the water ice band on inclination and ice fraction. For our best PDS 453 model, the depth of the \rev{water ice} band rises sharply between 76 and 80\degr\ as the column density increases, but it reaches a nearly flat plateau at $i\gtrsim81$\degr\ (edge-on disks). In other words, the \rev{water ice} band reaches saturation beyond a critical inclination, where it becomes impossible to quantify the amount of water ice. This phenomenon has already been reported in models by \citet{2005ApJ...622..463P} and \citet{2023A&A...677A..18S}, as well as in observations of the edge-on disk surrounding the L1527 protostar \citep{2012A&A...538A..57A}. While the critical inclination is dependent on the disk properties (mass, scale height, flaring, ...), this saturation effect will be relevant in the interpretation of upcoming JWST observations of highly inclined disks. \rev{Moreover, the depth of the water ice band evolves rapidly in the range of inclinations between $\sim$79\degr\ - 81\degr\, making it difficult to find a unique solution for the model.}
In addition, while the spectrum we used only covers the water ice band at 3.1\,$\mu$m, other ice species can be detected in edge-on disks \rev{also showing asymmetric \rev{water ice} bands \citep[e.g.,][]{2012A&A...538A..57A}. For instance, CO$_2$ presents two absorption \rev{ice} bands at 4.27\,$\mu$m and 15\,$\mu$m while CO features an ice band at 4.67\,$\mu$m}. We defer the \rev{modeling CO and CO$_2$ ices in prevision of future PDS 453 observations to a subsequent paper.}

This \rev{water ice} band saturation phenomenon is also observed as a function of the ice quantity (Fig. \ref{fig:ICE_SATURATION}). In our best model for PDS 453, saturation occurs once the water ice volume fraction reaches $\approx40\%$. \citet{2023A&A...679A.138S} also found that the H$_{2}$O, CO, and possibly CO$_{2}$ \rev{ice bands} are saturated in the HH\,48\,NE edge-on disk.

\subsection{Dust Properties}

\begin{center}
    \begin{figure}[h!]
        \centering
        \includegraphics[width=0.45\textwidth]{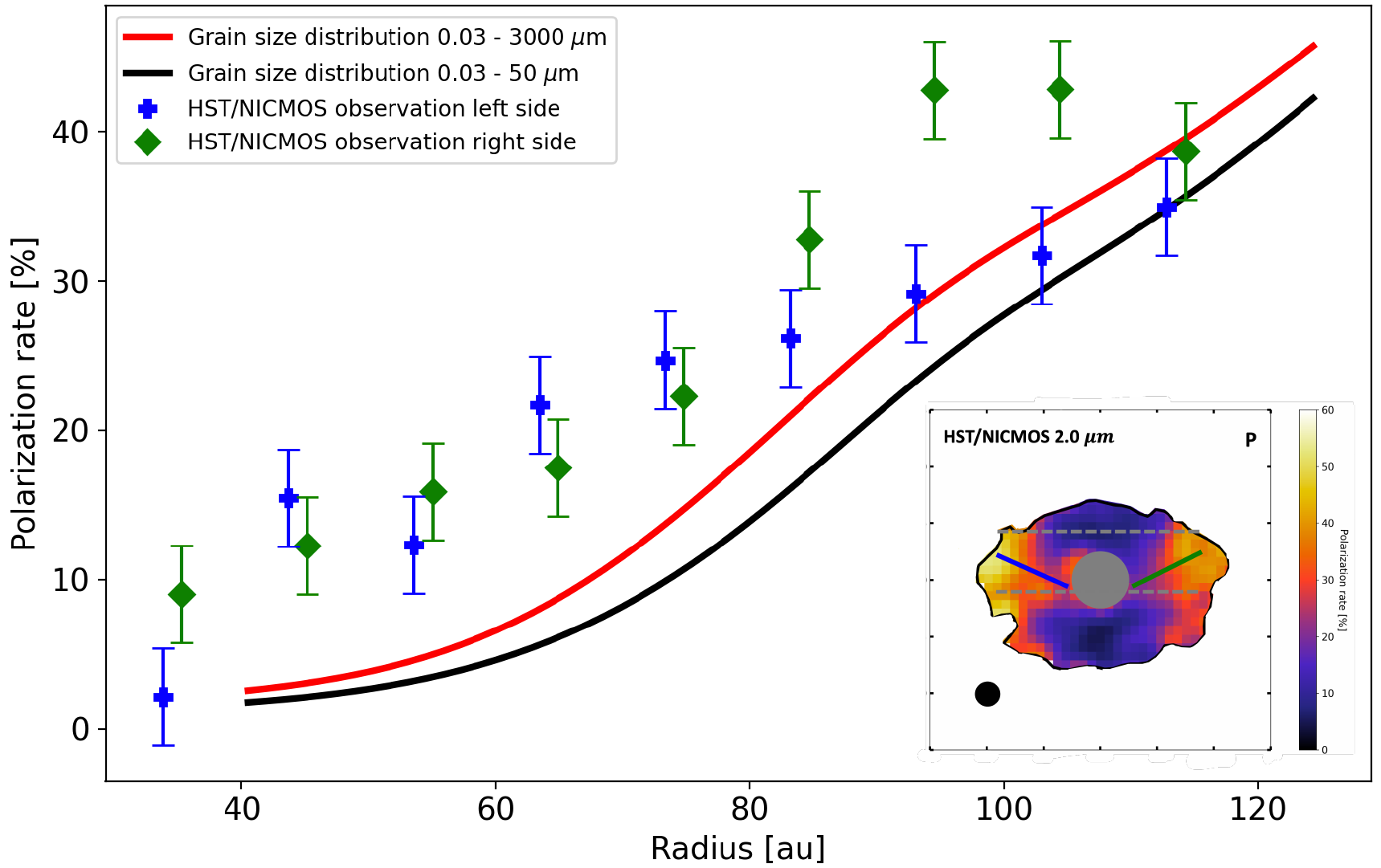}
        \caption{Observed (datapoints) and model (curves) polarization fraction measured along the top disk surface, from the edge of the coronagraphic mask to the outer edge of the disk. The red and black model curves only differ by the maximum grain size, with $a_{max}=3000$ and 50\,$\mu$m, respectively.
        The inset presents the same image as in Fig. \ref{fig:MODELS}, with blue and green solid lines corresponding to the radius along which the polarization rate is measured and the limits between the average pixel values representing by the two dashed gray lines.} 
        \label{fig:polar_rate_evolution}
    \end{figure}
\end{center}

In addition to the water ice \rev{band}, the polarization map across the disk provides information on the dust properties. Our best model broadly reproduces the trend of increasing polarization fraction as a function of projected distance from the star as well as the maximum polarization fraction of $\approx40\%$ (Fig. \ref{fig:polar_rate_evolution}). In an attempt to alter the polarization fraction profile, we computed another model in which the maximum grain size is set to 50\,$\mu$m, instead of 3mm. Since our model assumes fully mixed dust, reducing the maximum grain size represents a simplified approach to incorporating dust settling, as the millimeter grains are known to be concentrated in a thin midplane where they cannot scatter starlight \citep{villenave2020}. The resulting  polarization fraction throughout the disk is slightly lower than what is observed but the shape of the curve is unchanged. It is likely necessary to further reduce $a_{max}$ to obtain a better match. However, in fully mixed models, the long-wavelength end of the SED requires the presence of millimeter grains and the models are much worse overall matches to the observed properties of PDS 453. Models including settling would likely help solve this conundrum but are beyond the scope of this work.

\citet{2019ApJ...885...52T} suggest that a high polarization fraction of 65--75$\%$ points toward highly porous dust aggregates consisting of small monomer grains, whereas a lower polarization fraction of the order of 30$\%$ corresponds to more compact grains. Porosity, which we set at 50\% in our model, strongly affects the way that light is scattered off grains, in particular reducing back scattering. In addition, the degree of polarization can also be affected by the composition and the monomer radius within aggregates \citep{2022A&A...663A..57T}. 

Besides porosity, the shape of the grains also affects the polarization fraction. We adopted spherical homogeneous grains (Mie theory) for simplicity in this study. However, while little is known about the structure of dust particles in protoplanetary disks, studies of Solar System particles clearly point to their complex, aggregate nature \citep{Brownlee_1985, Bentley_2016, Mannel_2016}. \citet{Tazaki2023} have studied the disk around IM Lup and demonstrated that such complex morphology particles, with a fractal dimension of $\sim1.5$ and characteristic radius greater than 2\,$\mu$m are necessary to reproduce observations. Considering such particles in the case of PDS 453 may also help improve the match to observations.

\section{Conclusions}
\label{sec:conclusion}

We have presented new high-resolution near-infrared polarised intensity, total intensity, and linear polarization fraction images of the PDS 453 protoplanetary disk. We developed a radiative transfer model that reproduces adequately \rev{the geometry and size of the disk on the SPHERE image, the behavior of the polarization map from the NICMOS observation}, as well as the system SED and the depth of the 3.1\,$\mu$m water ice band. Our main conclusions are as follows:
\begin{itemize}
    \item[-]
The surface density profile of PDS 453 extends from near the stellar surface to 160\,au, with a sharp transition at 70\,au that produces a clear ring-like feature in the SPHERE image. Our model incorporates this as two distinct regions. The location of this transition roughly matches that of the water snow line in the disk upper surface.
    \item[-]
The PDS 453 disk is inclined at 80\degr\ which, in combination with the disk scale height, results in a rare configuration where our line of sight to the star grazes through the disk upper surface. The bright central point source that is observed in the system, which is much fainter than the star itself, is best interpreted as a forward scattering glint. This particular viewing geometry, in combination with a modest amount of foreground extinction, also accounts for the flat SED of the system from the optical to 100\,$\mu$m.
    \item[-]
PDS 453's disk is composed of a mixture of dust and water ice, accounting for the observed absorption \rev{band} at 3.1\,$\mu$m. In our model, the inner and outer contain 10 and 20\% by volume of water ice \rev{assuming Mie theory}. Our dust mixture also reproduces the fact that we partially detect the back side of the 70\,au ring in polarized intensity, as well as the
observed near-infrared polarization fraction.
    \item[-]
Exploring variations in inclination and water ice content from the best model for the PDS 453, we find that the central wavelength, spectral shape and depth of the 3.1\,$\mu$m water ice \rev{band} results from a balance between absorption and scattering. In particular, for disks seen at grazing angle (like PDS 453) or higher inclinations, we find that the depth of the \rev{water ice} band easily saturates when the column density through the line-of-sight becomes too high. This effect constitutes a barrier to the quantification of the water ice within highly inclined disks and  will have an impact on forthcoming edge-on disks observations by JWST. In general, it will only be possible to place a lower limit on the water ice content of these disks.
\end{itemize}

Observations from JWST and ALMA are expected to provide new images to constrain more parameter and thus improve the PDS 453 model presented here. From JWST spectrum, it will be possible to obtain ice profiles, notably CO and CO$_{2}$ which are currently unknown for this source. ALMA will notably help constraining the mass and the inclination of the disk, which are two crucial parameters to build more accurate models.

Understanding vertical structure through the study of edge-on disks, and the way grains grow, is crucial to our understanding of planetary formation mechanisms.

\begin{acknowledgements}
This project has received funding from the European Research Council (ERC) under the European Union’s Horizon Europe research and innovation program (grant agreement No. 101053020, project Dust2Planets, PI: F. M\'enard and grant agreement No. 101002188, project PROTOPLANETS, PI: M. Benisty). M.V., K.R.S., G.D., and S.G.W. acknowledge funding support from JWST GO program \#2562 provided by NASA through a grant from the Space Telescope Science Institute, which is operated by the Association of Universities for Research in Astronomy, Incorporated, under NASA contract NAS5-26555.
We are thankful to John P. Wisniewski for his help with the NICMOS data and discussion about the paper. We are thankful to Ga\"el Chauvin for the overall management of the SPHERE GTO survey.
SPHERE was designed and built by a consortium made of IPAG (Grenoble, France), MPIA (Heidelberg, Germany), LAM (Marseille, France), LESIA (Paris, France), Laboratoire Lagrange (Nice, France), INAF–Osservatorio di Padova (Italy), Observatoire de Gen\`eve (Switzerland), ETH Zurich (Switzerland), NOVA (Netherlands), ONERA (France) and ASTRON (Netherlands) in collaboration with ESO. SPHERE was funded by ESO, with additional contributions from CNRS (France), MPIA (Germany), INAF (Italy), FINES (Switzerland) and NOVA (Netherlands). 
\end{acknowledgements}

\bibliography{MyBibFMe}

\begin{thebibliography}{72}
\expandafter\ifx\csname natexlab\endcsname\relax\def\natexlab#1{#1}\fi

\bibitem[{{Aikawa} {et~al.}(2012){Aikawa}, {Kamuro}, {Sakon}, {Itoh}, {Terada}, {Noble}, {Pontoppidan}, {Fraser}, {Tamura}, {Kandori}, {Kawamura}, \& {Ueno}}]{2012A&A...538A..57A}
{Aikawa}, Y., {Kamuro}, D., {Sakon}, I., {et~al.} 2012, \aap, 538, A57

\bibitem[{{Andr{\'e}} {et~al.}(2010){Andr{\'e}}, {Men'shchikov}, {Bontemps}, {K{\"o}nyves}, {Motte}, {Schneider}, {Didelon}, {Minier}, {Saraceno}, {Ward-Thompson}, {di Francesco}, {White}, {Molinari}, {Testi}, {Abergel}, {Griffin}, {Henning}, {Royer}, {Mer{\'\i}n}, {Vavrek}, {Attard}, {Arzoumanian}, {Wilson}, {Ade}, {Aussel}, {Baluteau}, {Benedettini}, {Bernard}, {Blommaert}, {Cambr{\'e}sy}, {Cox}, {di Giorgio}, {Hargrave}, {Hennemann}, {Huang}, {Kirk}, {Krause}, {Launhardt}, {Leeks}, {Le Pennec}, {Li}, {Martin}, {Maury}, {Olofsson}, {Omont}, {Peretto}, {Pezzuto}, {Prusti}, {Roussel}, {Russeil}, {Sauvage}, {Sibthorpe}, {Sicilia-Aguilar}, {Spinoglio}, {Waelkens}, {Woodcraft}, \& {Zavagno}}]{andre2010}
{Andr{\'e}}, P., {Men'shchikov}, A., {Bontemps}, S., {et~al.} 2010, \aap, 518, L102

\bibitem[{{Andrews} {et~al.}(2018){Andrews}, {Huang}, {P{\'e}rez}, {Isella}, {Dullemond}, {Kurtovic}, {Guzm{\'a}n}, {Carpenter}, {Wilner}, {Zhang}, {Zhu}, {Birnstiel}, {Bai}, {Benisty}, {Hughes}, {{\"O}berg}, \& {Ricci}}]{Andrews2018}
{Andrews}, S.~M., {Huang}, J., {P{\'e}rez}, L.~M., {et~al.} 2018, \apjl, 869, L41

\bibitem[{{Avenhaus} {et~al.}(2018){Avenhaus}, {Quanz}, {Garufi}, {Perez}, {Casassus}, {Pinte}, {Bertrang}, {Caceres}, {Benisty}, \& {Dominik}}]{Avenhaus2018}
{Avenhaus}, H., {Quanz}, S.~P., {Garufi}, A., {et~al.} 2018, \apj, 863, 44

\bibitem[{{Ballering} {et~al.}(2021){Ballering}, {Cleeves}, \& {Anderson}}]{Ballering_2021}
{Ballering}, N.~P., {Cleeves}, L.~I., \& {Anderson}, D.~E. 2021, \apj, 920, 115

\bibitem[{{Baraffe} {et~al.}(1998){Baraffe}, {Chabrier}, {Allard}, \& {Hauschildt}}]{Baraffe_1998}
{Baraffe}, I., {Chabrier}, G., {Allard}, F., \& {Hauschildt}, P.~H. 1998, \aap, 337, 403

\bibitem[{Benisty {et~al.}(2022)Benisty, Dominik, Follette, Garufi, Ginski, Hashimoto, Keppler, Kley, \& Monnier}]{benisty2022optical}
Benisty, M., Dominik, C., Follette, K., {et~al.} 2022, Optical and Near-infrared View of Planet-forming Disks and Protoplanets

\bibitem[{{Benisty} {et~al.}(2023){Benisty}, {Dominik}, {Follette}, {Garufi}, {Ginski}, {Hashimoto}, {Keppler}, {Kley}, \& {Monnier}}]{benisty2023}
{Benisty}, M., {Dominik}, C., {Follette}, K., {et~al.} 2023, in Astronomical Society of the Pacific Conference Series, Vol. 534, Protostars and Planets VII, ed. S.~{Inutsuka}, Y.~{Aikawa}, T.~{Muto}, K.~{Tomida}, \& M.~{Tamura}, 605

\bibitem[{{Bentley} {et~al.}(2016){Bentley}, {Schmied}, {Mannel}, {Torkar}, {Jeszenszky}, {Romstedt}, {Levasseur-Regourd}, {Weber}, {Jessberger}, {Ehrenfreund}, {Koeberl}, \& {Havnes}}]{Bentley_2016}
{Bentley}, M.~S., {Schmied}, R., {Mannel}, T., {et~al.} 2016, \nat, 537, 73

\bibitem[{{Beuzit} {et~al.}(2019){Beuzit}, {Vigan}, {Mouillet}, {Dohlen}, {Gratton}, {Boccaletti}, {Sauvage}, {Schmid}, {Langlois}, {Petit}, {Baruffolo}, {Feldt}, {Milli}, {Wahhaj}, {Abe}, {Anselmi}, {Antichi}, {Barette}, {Baudrand}, {Baudoz}, {Bazzon}, {Bernardi}, {Blanchard}, {Brast}, {Bruno}, {Buey}, {Carbillet}, {Carle}, {Cascone}, {Chapron}, {Charton}, {Chauvin}, {Claudi}, {Costille}, {De Caprio}, {de Boer}, {Delboulb{\'e}}, {Desidera}, {Dominik}, {Downing}, {Dupuis}, {Fabron}, {Fantinel}, {Farisato}, {Feautrier}, {Fedrigo}, {Fusco}, {Gigan}, {Ginski}, {Girard}, {Giro}, {Gisler}, {Gluck}, {Gry}, {Henning}, {Hubin}, {Hugot}, {Incorvaia}, {Jaquet}, {Kasper}, {Lagadec}, {Lagrange}, {Le Coroller}, {Le Mignant}, {Le Ruyet}, {Lessio}, {Lizon}, {Llored}, {Lundin}, {Madec}, {Magnard}, {Marteaud}, {Martinez}, {Maurel}, {M{\'e}nard}, {Mesa}, {M{\"o}ller-Nilsson}, {Moulin}, {Moutou}, {Orign{\'e}}, {Parisot}, {Pavlov}, {Perret}, {Pragt}, {Puget}, {Rabou}, {Ramos}, {Reess}, {Rigal}, {Rochat}, {Roelfsema}, {Rousset},
  {Roux}, {Saisse}, {Salasnich}, {Santambrogio}, {Scuderi}, {Segransan}, {Sevin}, {Siebenmorgen}, {Soenke}, {Stadler}, {Suarez}, {Tiph{\`e}ne}, {Turatto}, {Udry}, {Vakili}, {Waters}, {Weber}, {Wildi}, {Zins}, \& {Zurlo}}]{Beuzit2019}
{Beuzit}, J.~L., {Vigan}, A., {Mouillet}, D., {et~al.} 2019, \aap, 631, A155

\bibitem[{{Birnstiel} {et~al.}(2018){Birnstiel}, {Dullemond}, {Zhu}, {Andrews}, {Bai}, {Wilner}, {Carpenter}, {Huang}, {Isella}, {Benisty}, {P{\'e}rez}, \& {Zhang}}]{Birnstiel2018}
{Birnstiel}, T., {Dullemond}, C.~P., {Zhu}, Z., {et~al.} 2018, \apjl, 869, L45

\bibitem[{{Boogert} {et~al.}(2015){Boogert}, {Gerakines}, \& {Whittet}}]{2015ARA&A..53..541B}
{Boogert}, A.~C.~A., {Gerakines}, P.~A., \& {Whittet}, D. C.~B. 2015, \araa, 53, 541

\bibitem[{{Brownlee}(1985)}]{Brownlee_1985}
{Brownlee}, D.~E. 1985, Annual Review of Earth and Planetary Sciences, 13, 147

\bibitem[{{Burrows} {et~al.}(1996){Burrows}, {Stapelfeldt}, {Watson}, {Krist}, {Ballester}, {Clarke}, {Crisp}, {Gallagher}, {Griffiths}, {Hester}, {Hoessel}, {Holtzman}, {Mould}, {Scowen}, {Trauger}, \& {Westphal}}]{Burrows1996}
{Burrows}, C.~J., {Stapelfeldt}, K.~R., {Watson}, A.~M., {et~al.} 1996, \apj, 473, 437

\bibitem[{{D'Alessio} {et~al.}(2001){D'Alessio}, {Calvet}, \& {Hartmann}}]{Dalessio_2001}
{D'Alessio}, P., {Calvet}, N., \& {Hartmann}, L. 2001, \apj, 553, 321

\bibitem[{{Dartois} {et~al.}(2024){Dartois}, {Noble}, {Caselli}, {Fraser}, {Jim{\'e}nez-Serra}, {Mat{\'e}}, {McClure}, {Melnick}, {Pendleton}, {Shimonishi}, {Smith}, {Sturm}, {Taillard}, {Wakelam}, {Boogert}, {Drozdovskaya}, {Erkal}, {Harsono}, {Herrero}, {Ioppolo}, {Linnartz}, {McGuire}, {Perotti}, {Qasim}, \& {Rocha}}]{2024NatAs.tmp....6D}
{Dartois}, E., {Noble}, J.~A., {Caselli}, P., {et~al.} 2024, Nature Astronomy

\bibitem[{{de Boer} {et~al.}(2020){de Boer}, {Langlois}, {van Holstein}, {Girard}, {Mouillet}, {Vigan}, {Dohlen}, {Snik}, {Keller}, {Ginski}, {Stam}, {Milli}, {Wahhaj}, {Kasper}, {Schmid}, {Rabou}, {Gluck}, {Hugot}, {Perret}, {Martinez}, {Weber}, {Pragt}, {Sauvage}, {Boccaletti}, {Le Coroller}, {Dominik}, {Henning}, {Lagadec}, {M{\'e}nard}, {Turatto}, {Udry}, {Chauvin}, {Feldt}, \& {Beuzit}}]{deBoer2020}
{de Boer}, J., {Langlois}, M., {van Holstein}, R.~G., {et~al.} 2020, \aap, 633, A63

\bibitem[{{Derkink} {et~al.}(2024){Derkink}, {Ginski}, {Pinilla}, {Kurtovic}, {Kaper}, {de Koter}, {Valeg{\r{a}}rd}, {Mamajek}, {Backs}, {Benisty}, {Birnstiel}, {Columba}, {Dominik}, {Garufi}, {Hogerheijde}, {van Holstein}, {Huang}, {M{\'e}nard}, {Rab}, {Ram{\'\i}rez-Tannus}, {Ribas}, {Williams}, \& {Zurlo}}]{Derkink2024}
{Derkink}, A., {Ginski}, C., {Pinilla}, P., {et~al.} 2024, \aap, 688, A149

\bibitem[{{Dohlen} {et~al.}(2008){Dohlen}, {Langlois}, {Saisse}, {Hill}, {Origne}, {Jacquet}, {Fabron}, {Blanc}, {Llored}, {Carle}, {Moutou}, {Vigan}, {Boccaletti}, {Carbillet}, {Mouillet}, \& {Beuzit}}]{Dohlen2008}
{Dohlen}, K., {Langlois}, M., {Saisse}, M., {et~al.} 2008, Society of Photo-Optical Instrumentation Engineers (SPIE) Conference Series, Vol. 7014, {The infra-red dual imaging and spectrograph for SPHERE: design and performance} (SPIE), 70143L

\bibitem[{{Duch{\^e}ne} {et~al.}(2024){Duch{\^e}ne}, {M{\'e}nard}, {Stapelfeldt}, {Villenave}, {Wolff}, {Perrin}, {Pinte}, {Tazaki}, \& {Padgett}}]{Duchene2024}
{Duch{\^e}ne}, G., {M{\'e}nard}, F., {Stapelfeldt}, K.~R., {et~al.} 2024, \aj, 167, 77

\bibitem[{{Esplin} \& {Luhman}(2020)}]{Esplin_2020}
{Esplin}, T.~L. \& {Luhman}, K.~L. 2020, \aj, 159, 282

\bibitem[{{Garufi} {et~al.}(2020){Garufi}, {Avenhaus}, {P{\'e}rez}, {Quanz}, {van Holstein}, {Bertrang}, {Casassus}, {Cieza}, {Principe}, {van der Plas}, \& {Zurlo}}]{2020A&A...633A..82G}
{Garufi}, A., {Avenhaus}, H., {P{\'e}rez}, S., {et~al.} 2020, \aap, 633, A82

\bibitem[{{Garufi} {et~al.}(2024){Garufi}, {Ginski}, {van Holstein}, {Benisty}, {Manara}, {P{\'e}rez}, {Pinilla}, {Ribas}, {Weber}, {Williams}, {Cieza}, {Dominik}, {Facchini}, {Huang}, {Zurlo}, {Bae}, {Hagelberg}, {Henning}, {Hogerheijde}, {Janson}, {M{\'e}nard}, {Messina}, {Meyer}, {Pinte}, {Quanz}, {Rigliaco}, {Roccatagliata}, {Schmid}, {Szul{\'a}gyi}, {van Boekel}, {Wahhaj}, {Antichi}, {Baruffolo}, \& {Moulin}}]{2024A&A...685A..53G}
{Garufi}, A., {Ginski}, C., {van Holstein}, R.~G., {et~al.} 2024, \aap, 685, A53

\bibitem[{{Glauser} {et~al.}(2008){Glauser}, {M{\'e}nard}, {Pinte}, {Duch{\^e}ne}, {G{\"u}del}, {Monin}, \& {Padgett}}]{Glauser2008}
{Glauser}, A.~M., {M{\'e}nard}, F., {Pinte}, C., {et~al.} 2008, \aap, 485, 531

\bibitem[{{Greene} {et~al.}(1994){Greene}, {Wilking}, {Andre}, {Young}, \& {Lada}}]{2008A&A94}
{Greene}, T.~P., {Wilking}, B.~A., {Andre}, P., {Young}, E.~T., \& {Lada}, C.~J. 1994, \apj, 434, 614

\bibitem[{{Hines} {et~al.}(2000){Hines}, {Schmidt}, \& {Schneider}}]{2000AAS...197.1210H}
{Hines}, D.~C., {Schmidt}, G.~D., \& {Schneider}, G. 2000, in American Astronomical Society Meeting Abstracts, Vol. 197, American Astronomical Society Meeting Abstracts, 12.10

\bibitem[{{Hollenbach} {et~al.}(2009){Hollenbach}, {Kaufman}, {Bergin}, \& {Melnick}}]{Hollenbach_2009}
{Hollenbach}, D., {Kaufman}, M.~J., {Bergin}, E.~A., \& {Melnick}, G.~J. 2009, \apj, 690, 1497

\bibitem[{{Jennings} {et~al.}(2022){Jennings}, {Tazzari}, {Clarke}, {Booth}, \& {Rosotti}}]{Jennings_2022}
{Jennings}, J., {Tazzari}, M., {Clarke}, C.~J., {Booth}, R.~A., \& {Rosotti}, G.~P. 2022, \mnras, 514, 6053

\bibitem[{{Johansen} {et~al.}(2014){Johansen}, {Blum}, {Tanaka}, {Ormel}, {Bizzarro}, \& {Rickman}}]{Johansen_2014}
{Johansen}, A., {Blum}, J., {Tanaka}, H., {et~al.} 2014, in Protostars and Planets VI, ed. H.~{Beuther}, R.~S. {Klessen}, C.~P. {Dullemond}, \& T.~{Henning}, 547--570

\bibitem[{{Kaeufer} {et~al.}(2023){Kaeufer}, {Woitke}, {Min}, {Kamp}, \& {Pinte}}]{2023A&A...672A..30K}
{Kaeufer}, T., {Woitke}, P., {Min}, M., {Kamp}, I., \& {Pinte}, C. 2023, \aap, 672, A30

\bibitem[{{Keppler} {et~al.}(2018){Keppler}, {Benisty}, {M{\"u}ller}, {Henning}, {van Boekel}, {Cantalloube}, {Ginski}, {van Holstein}, {Maire}, {Pohl}, {Samland }, {Avenhaus}, {Baudino}, {Boccaletti}, {de Boer}, {Bonnefoy}, {Chauvin}, {Desidera}, {Langlois}, {Lazzoni}, {Marleau}, {Mordasini}, {Pawellek}, {Stolker}, {Vigan}, {Zurlo}, {Birnstiel}, {Brandner}, {Feldt}, {Flock}, {Girard}, {Gratton}, {Hagelberg}, {Isella}, {Janson}, {Juhasz}, {Kemmer}, {Kral}, {Lagrange}, {Launhardt}, {Matter}, {M{\'e}nard}, {Milli}, {Molli{\`e}re}, {Olofsson}, {P{\'e}rez}, {Pinilla}, {Pinte}, {Quanz}, {Schmidt}, {Udry}, {Wahhaj}, {Williams}, {Buenzli}, {Cudel}, {Dominik}, {Galicher}, {Kasper}, {Lannier}, {Mesa}, {Mouillet}, {Peretti}, {Perrot}, {Salter}, {Sissa}, {Wildi}, {Abe}, {Antichi}, {Augereau}, {Baruffolo}, {Baudoz}, {Bazzon}, {Beuzit}, {Blanchard}, {Brems}, {Buey}, {De Caprio}, {Carbillet}, {Carle}, {Cascone}, {Cheetham}, {Claudi}, {Costille}, {Delboulb{\'e}}, {Dohlen}, {Fantinel}, {Feautrier}, {Fusco}, {Giro}, {Gluck},
  {Gry}, {Hubin}, {Hugot}, {Jaquet}, {Le Mignant}, {Llored}, {Madec}, {Magnard}, {Martinez}, {Maurel}, {Meyer}, {M{\"o}ller-Nilsson}, {Moulin}, {Mugnier}, {Orign{\'e}}, {Pavlov}, {Perret}, {Petit}, {Pragt}, {Puget}, {Rabou}, {Ramos}, {Rigal}, {Rochat}, {Roelfsema}, {Rousset}, {Roux}, {Salasnich}, {Sauvage}, {Sevin}, {Soenke}, {Stadler}, {Suarez}, {Turatto}, \& {Weber}}]{Keppler2018}
{Keppler}, M., {Benisty}, M., {M{\"u}ller}, A., {et~al.} 2018, \aap, 617, A44

\bibitem[{{Leger} {et~al.}(1983){Leger}, {Gauthier}, {Defourneau}, \& {Rouan}}]{1983A&A...117..164L}
{Leger}, A., {Gauthier}, S., {Defourneau}, D., \& {Rouan}, D. 1983, \aap, 117, 164

\bibitem[{{Li} \& {Greenberg}(1998)}]{1998A&A...331..291L}
{Li}, A. \& {Greenberg}, J.~M. 1998, \aap, 331, 291

\bibitem[{{Long} {et~al.}(2018){Long}, {Pinilla}, {Herczeg}, {Harsono}, {Dipierro}, {Pascucci}, {Hendler}, {Tazzari}, {Ragusa}, {Salyk}, {Edwards}, {Lodato}, {van de Plas}, {Johnstone}, {Liu}, {Boehler}, {Cabrit}, {Manara}, {Menard}, {Mulders}, {Nisini}, {Fischer}, {Rigliaco}, {Banzatti}, {Avenhaus}, \& {Gully-Santiago}}]{Long2018}
{Long}, F., {Pinilla}, P., {Herczeg}, G.~J., {et~al.} 2018, \apj, 869, 17

\bibitem[{{Maire} {et~al.}(2016){Maire}, {Langlois}, {Dohlen}, {Lagrange}, {Gratton}, {Chauvin}, {Desidera}, {Girard}, {Milli}, {Vigan}, {Zins}, {Delorme}, {Beuzit}, {Claudi}, {Feldt}, {Mouillet}, {Puget}, {Turatto}, \& {Wildi}}]{Maire2016}
{Maire}, A.-L., {Langlois}, M., {Dohlen}, K., {et~al.} 2016, in Society of Photo-Optical Instrumentation Engineers (SPIE) Conference Series, Vol. 9908, Ground-based and Airborne Instrumentation for Astronomy VI, 990834

\bibitem[{{Mannel} {et~al.}(2016){Mannel}, {Bentley}, {Schmied}, {Jeszenszky}, {Levasseur-Regourd}, {Romstedt}, \& {Torkar}}]{Mannel_2016}
{Mannel}, T., {Bentley}, M.~S., {Schmied}, R., {et~al.} 2016, \mnras, 462, S304

\bibitem[{{Mathis} \& {Whiffen}(1989)}]{1989ApJ...341..808M}
{Mathis}, J.~S. \& {Whiffen}, G. 1989, \apj, 341, 808

\bibitem[{{McClure} {et~al.}(2023){McClure}, {Rocha}, {Pontoppidan}, {Crouzet}, {Chu}, {Dartois}, {Lamberts}, {Noble}, {Pendleton}, {Perotti}, {Qasim}, {Rachid}, {Smith}, {Sun}, {Beck}, {Boogert}, {Brown}, {Caselli}, {Charnley}, {Cuppen}, {Dickinson}, {Drozdovskaya}, {Egami}, {Erkal}, {Fraser}, {Garrod}, {Harsono}, {Ioppolo}, {Jim{\'e}nez-Serra}, {Jin}, {J{\o}rgensen}, {Kristensen}, {Lis}, {McCoustra}, {McGuire}, {Melnick}, {{\~A}-berg}, {Palumbo}, {Shimonishi}, {Sturm}, {van Dishoeck}, \& {Linnartz}}]{2023NatAs...7..431M}
{McClure}, M.~K., {Rocha}, W.~R.~M., {Pontoppidan}, K.~M., {et~al.} 2023, Nature Astronomy, 7, 431

\bibitem[{{Min} {et~al.}(2012){Min}, {Canovas}, {Mulders}, \& {Keller}}]{2012A&A...537A..75M}
{Min}, M., {Canovas}, H., {Mulders}, G.~D., \& {Keller}, C.~U. 2012, \aap, 537, A75

\bibitem[{{Murakawa}(2010)}]{2010A&A...518A..63M}
{Murakawa}, K. 2010, \aap, 518, A63

\bibitem[{{Oka} {et~al.}(2011){Oka}, {Nakamoto}, \& {Ida}}]{Oka_2011}
{Oka}, A., {Nakamoto}, T., \& {Ida}, S. 2011, \apj, 738, 141

\bibitem[{Pecaut \& Mamajek(2013)}]{Pecaut_2013}
Pecaut, M.~J. \& Mamajek, E.~E. 2013, \apjs, 208, 9

\bibitem[{{Pecaut} {et~al.}(2012){Pecaut}, {Mamajek}, \& {Bubar}}]{Pecaut_2012}
{Pecaut}, M.~J., {Mamajek}, E.~E., \& {Bubar}, E.~J. 2012, \apj, 746, 154

\bibitem[{{Perrin}(2006)}]{Perrin2006PhDT.......203P}
{Perrin}, M.~D. 2006, PhD thesis, University of California, Berkeley

\bibitem[{{Perrin} {et~al.}(2009){Perrin}, {Schneider}, {Duchene}, {Pinte}, {Grady}, {Wisniewski}, \& {Hines}}]{2009ApJ...707L.132P}
{Perrin}, M.~D., {Schneider}, G., {Duchene}, G., {et~al.} 2009, \apjl, 707, L132

\bibitem[{{Perrin} {et~al.}(2010){Perrin}, {Schnieder}, {Duchene}, {Hines}, {Pinte}, {Fitzgerald}, {Wisniewski}, \& {HST GO 11155 Team}}]{2010AAS...21542812P}
{Perrin}, M.~D., {Schnieder}, G., {Duchene}, G., {et~al.} 2010, in American Astronomical Society Meeting Abstracts, Vol. 215, American Astronomical Society Meeting Abstracts \#215, 428.12

\bibitem[{{Pinte} {et~al.}(2006){Pinte}, {M{\'e}nard}, {Duch{\^e}ne}, \& {Bastien}}]{Pinte2006}
{Pinte}, C., {M{\'e}nard}, F., {Duch{\^e}ne}, G., \& {Bastien}, P. 2006, \aap, 459, 797

\bibitem[{{Pinte} {et~al.}(2008){Pinte}, {Padgett}, {M{\'e}nard}, {Stapelfeldt}, {Schneider}, {Olofsson}, {Pani{\'c}}, {Augereau}, {Duch{\^e}ne}, {Krist}, {Pontoppidan}, {Perrin}, {Grady}, {Kessler-Silacci}, {van Dishoeck}, {Lommen}, {Silverstone}, {Hines}, {Wolf}, {Blake}, {Henning}, \& {Stecklum}}]{Pinte2008}
{Pinte}, C., {Padgett}, D.~L., {M{\'e}nard}, F., {et~al.} 2008, \aap, 489, 633

\bibitem[{Pollack {et~al.}(1994)Pollack, Hollenbach, Beckwith, Simonelli, Roush, \& Fong}]{Pollack_1994}
Pollack, J., Hollenbach, D., Beckwith, S., {et~al.} 1994, The Astrophysical Journal, 421, 615

\bibitem[{{Pontoppidan} {et~al.}(2005){Pontoppidan}, {Dullemond}, {van Dishoeck}, {Blake}, {Boogert}, {Evans}, {Kessler-Silacci}, \& {Lahuis}}]{2005ApJ...622..463P}
{Pontoppidan}, K.~M., {Dullemond}, C.~P., {van Dishoeck}, E.~F., {et~al.} 2005, \apj, 622, 463

\bibitem[{Preibisch \& Mamajek(2008)}]{preibisch2008nearest}
Preibisch, T. \& Mamajek, E. 2008, The Nearest OB Association: Scorpius-Centaurus (Sco OB2)

\bibitem[{{Ratzenb{\"o}ck} {et~al.}(2023){Ratzenb{\"o}ck}, {Gro{\ss}schedl}, {Alves}, {Miret-Roig}, {Bomze}, {Forbes}, {Goodman}, {Hacar}, {Lin}, {Meingast}, {M{\"o}ller}, {Piecka}, {Posch}, {Rottensteiner}, {Swiggum}, \& {Zucker}}]{ratzenbock23}
{Ratzenb{\"o}ck}, S., {Gro{\ss}schedl}, J.~E., {Alves}, J., {et~al.} 2023, \aap, 678, A71

\bibitem[{Sartori {et~al.}(2010)Sartori, Gregorio-Hetem, Rodrigues, Hetem, \& Batalha}]{Sartori_2010}
Sartori, M.~J., Gregorio-Hetem, J., Rodrigues, C.~V., Hetem, A., \& Batalha, C. 2010, The Astronomical Journal, 139, 27

\bibitem[{{Schneider} {et~al.}(2005){Schneider}, {Silverstone}, \& {Hines}}]{2005ApJ...629L.117S}
{Schneider}, G., {Silverstone}, M.~D., \& {Hines}, D.~C. 2005, \apjl, 629, L117

\bibitem[{{Sturm} {et~al.}(2023{\natexlab{a}}){Sturm}, {McClure}, {Beck}, {Harsono}, {Bergner}, {Dartois}, {Boogert}, {Chiar}, {Cordiner}, {Drozdovskaya}, {Ioppolo}, {Law}, {Linnartz}, {Lis}, {Melnick}, {McGuire}, {Noble}, {{\"O}berg}, {Palumbo}, {Pendleton}, {Perotti}, {Pontoppidan}, {Qasim}, {Rocha}, {Terada}, {Urso}, \& {van Dishoeck}}]{2023A&A...679A.138S}
{Sturm}, J.~A., {McClure}, M.~K., {Beck}, T.~L., {et~al.} 2023{\natexlab{a}}, \aap, 679, A138

\bibitem[{{Sturm} {et~al.}(2023{\natexlab{b}}){Sturm}, {McClure}, {Bergner}, {Harsono}, {Dartois}, {Drozdovskaya}, {Ioppolo}, {{\"O}berg}, {Law}, {Palumbo}, {Pendleton}, {Rocha}, {Terada}, \& {Urso}}]{2023A&A...677A..18S}
{Sturm}, J.~A., {McClure}, M.~K., {Bergner}, J.~B., {et~al.} 2023{\natexlab{b}}, \aap, 677, A18

\bibitem[{{Sturm} {et~al.}(2024){Sturm}, {McClure}, {Harsono}, {Bergner}, {Dartois}, {Boogert}, {Cordiner}, {Drozdovskaya}, {Ioppolo}, {Law}, {Lis}, {McGuire}, {Melnick}, {Noble}, {{\"O}berg}, {Palumbo}, {Pendleton}, {Perotti}, {Rocha}, {Urso}, \& {van Dishoeck}}]{Sturm_2024}
{Sturm}, J.~A., {McClure}, M.~K., {Harsono}, D., {et~al.} 2024, \aap, 689, A92

\bibitem[{{Sturm} {et~al.}(2023{\natexlab{c}}){Sturm}, {McClure}, {Law}, {Harsono}, {Bergner}, {Dartois}, {Drozdovskaya}, {Ioppolo}, {{\"O}berg}, {Palumbo}, {Pendleton}, {Rocha}, {Terada}, \& {Urso}}]{Sturm2023}
{Sturm}, J.~A., {McClure}, M.~K., {Law}, C.~J., {et~al.} 2023{\natexlab{c}}, \aap, 677, A17

\bibitem[{{Tazaki} \& {Dominik}(2022)}]{2022A&A...663A..57T}
{Tazaki}, R. \& {Dominik}, C. 2022, \aap, 663, A57

\bibitem[{{Tazaki} {et~al.}(2023){Tazaki}, {Ginski}, \& {Dominik}}]{Tazaki2023}
{Tazaki}, R., {Ginski}, C., \& {Dominik}, C. 2023, \apjl, 944, L43

\bibitem[{{Tazaki} {et~al.}(2021){Tazaki}, {Murakawa}, {Muto}, {Honda}, \& {Inoue}}]{2021ApJ...921..173T}
{Tazaki}, R., {Murakawa}, K., {Muto}, T., {Honda}, M., \& {Inoue}, A.~K. 2021, \apj, 921, 173

\bibitem[{{Tazaki} {et~al.}(2019){Tazaki}, {Tanaka}, {Kataoka}, {Okuzumi}, \& {Muto}}]{2019ApJ...885...52T}
{Tazaki}, R., {Tanaka}, H., {Kataoka}, A., {Okuzumi}, S., \& {Muto}, T. 2019, \apj, 885, 52

\bibitem[{{Terada} \& {Tokunaga}(2017)}]{Terada2017}
{Terada}, H. \& {Tokunaga}, A.~T. 2017, \apj, 834, 115

\bibitem[{Terada \& Tokunaga(2017)}]{Terada_2017}
Terada, H. \& Tokunaga, A.~T. 2017, The Astrophysical Journal, 834, 115

\bibitem[{{Valeg{\r{a}}rd} {et~al.}(2024){Valeg{\r{a}}rd}, {Ginski}, {Derkink}, {Garufi}, {Dominik}, {Ribas}, {Williams}, {Benisty}, {Birnstiel}, {Facchini}, {Columba}, {Hogerheijde}, {van Holstein}, {Huang}, {Kenworthy}, {Manara}, {Pinilla}, {Rab}, {Sulaiman}, \& {Zurlo}}]{2024A&A...685A..54V}
{Valeg{\r{a}}rd}, P.~G., {Ginski}, C., {Derkink}, A., {et~al.} 2024, \aap, 685, A54

\bibitem[{{van Holstein} {et~al.}(2020){van Holstein}, {Girard}, {de Boer}, {Snik}, {Milli}, {Stam}, {Ginski}, {Mouillet}, {Wahhaj}, {Schmid}, {Keller}, {Langlois}, {Dohlen}, {Vigan}, {Pohl}, {Carbillet}, {Fantinel}, {Maurel}, {Orign{\'e}}, {Petit}, {Ramos}, {Rigal}, {Sevin}, {Boccaletti}, {Le Coroller}, {Dominik}, {Henning}, {Lagadec}, {M{\'e}nard}, {Turatto}, {Udry}, {Chauvin}, {Feldt}, \& {Beuzit}}]{vanHolstein2020}
{van Holstein}, R.~G., {Girard}, J.~H., {de Boer}, J., {et~al.} 2020, \aap, 633, A64

\bibitem[{{Villenave} {et~al.}(2019){Villenave}, {Benisty}, {Dent}, {M{\'e}nard}, {Garufi}, {Ginski}, {Pinilla}, {Pinte}, {Williams}, {de Boer}, {Morino}, {Fukagawa}, {Dominik}, {Flock}, {Henning}, {Juh{\'a}sz}, {Keppler}, {Muro-Arena}, {Olofsson}, {P{\'e}rez}, {van der Plas}, {Zurlo}, {Carle}, {Feautrier}, {Pavlov}, {Pragt}, {Ramos}, {Sauvage}, {Stadler}, \& {Weber}}]{Villenave_2019}
{Villenave}, M., {Benisty}, M., {Dent}, W.~R.~F., {et~al.} 2019, \aap, 624, A7

\bibitem[{{Villenave} {et~al.}(2020){Villenave}, {M{\'e}nard}, {Dent}, {Duch{\^e}ne}, {Stapelfeldt}, {Benisty}, {Boehler}, {van der Plas}, {Pinte}, {Telkamp}, {Wolff}, {Flores}, {Lesur}, {Louvet}, {Riols}, {Dougados}, {Williams}, \& {Padgett}}]{villenave2020}
{Villenave}, M., {M{\'e}nard}, F., {Dent}, W.~R.~F., {et~al.} 2020, \aap, 642, A164

\bibitem[{{Villenave} {et~al.}(2024){Villenave}, {Stapelfeldt}, {Duch{\^e}ne}, {M{\'e}nard}, {Wolff}, {Perrin}, {Pinte}, {Tazaki}, \& {Padgett}}]{Villenave2024}
{Villenave}, M., {Stapelfeldt}, K.~R., {Duch{\^e}ne}, G., {et~al.} 2024, \apj, 961, 95

\bibitem[{{Vioque, M.} {et~al.}(2018){Vioque, M.}, {Oudmaijer, R. D.}, {Baines, D.}, {Mendigut\'{\i}a, I.}, \& {P\'erez-Mart\'{\i}nez, R.}}]{Vioque_2018}
{Vioque, M.}, {Oudmaijer, R. D.}, {Baines, D.}, {Mendigut\'{\i}a, I.}, \& {P\'erez-Mart\'{\i}nez, R.} 2018, A\&A, 620, A128

\bibitem[{{Watson} \& {Stapelfeldt}(2007)}]{Watson2007}
{Watson}, A.~M. \& {Stapelfeldt}, K.~R. 2007, \aj, 133, 845

\bibitem[{{Woitke} {et~al.}(2019){Woitke}, {Kamp}, {Antonellini}, {Anthonioz}, {Baldovin-Saveedra}, {Carmona}, {Dionatos}, {Dominik}, {Greaves}, {G{\"u}del}, {Ilee}, {Liebhardt}, {Menard}, {Min}, {Pinte}, {Rab}, {Rigon}, {Thi}, {Thureau}, \& {Waters}}]{2019PASP..131f4301W}
{Woitke}, P., {Kamp}, I., {Antonellini}, S., {et~al.} 2019, \pasp, 131, 064301

\end{thebibliography}

\onecolumn
\begin{appendix}

\section{HST/NICMOS Total Intensity and Polarization Intensity Image at 2\,$\mu$m}

\begin{center}
    \begin{figure}[h]
        \centering
        \includegraphics[width=1\textwidth]{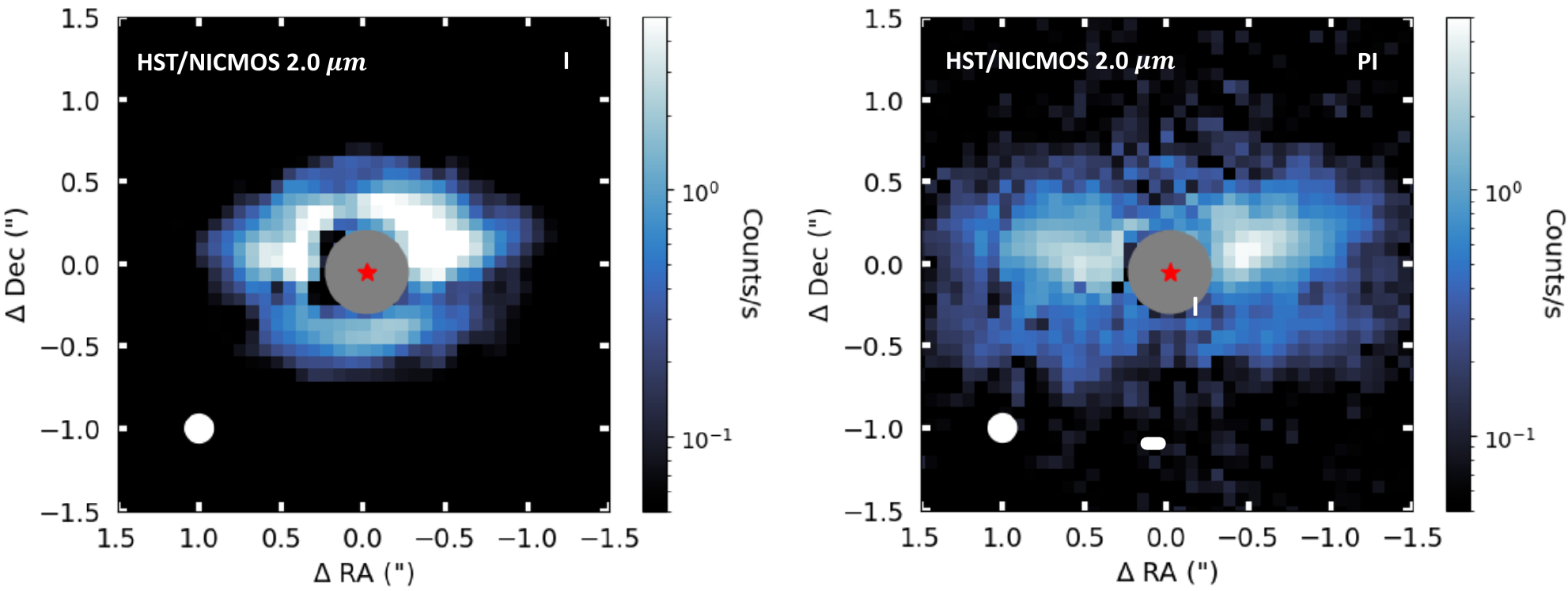}
        \caption{
        \textbf{\textit{Left panel:}} HST/NICMOS scattered light total intensity image at 2\,$\mu$m shown using a logarithmic stretch.
        \textbf{\textit{Right panel:}} HST/NICMOS scattered light polarized intensity image 2\,$\mu$m shown using a logarithmic stretch. \final{The red star symbol corresponds to the position of the bright central point source.}}
        \label{fig:hst_2_micron}
    \end{figure}
\end{center}

Figure \ref{fig:hst_2_micron} presents the 2\,$\mu$m HST/NICMOS scattered light total intensity and polarized intensity images.

\section{Single-zone model}

\begin{center}
    \begin{figure}[h]
        \centering
        \includegraphics[width=1\textwidth]{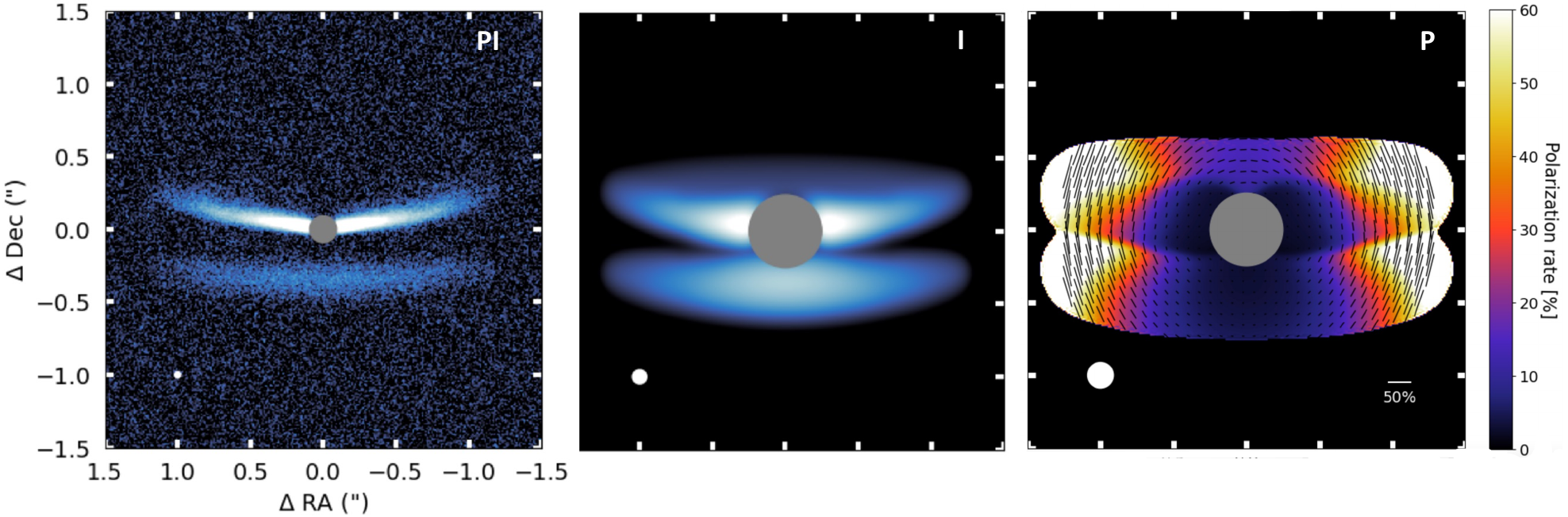}
        \caption{
        Synthetic observations of our PDS 453 model with a single zone, with similar scales and stretches as in Fig.\,\ref{fig:OBSERVATIONS} and Fig.\,\ref{fig:MODELS}. All images have been convolved by the appropriate PSF after subtracting the central point source to mimic the effect of the coronagraph.
       \textbf{\textit{Left panel:}} Scattered light 1.6\,$\mu$m polarized intensity image shown a the SPHERE resolution. \textbf{\textit{Middle panel:}} Scattered light F110W total intensity image at the NICMOS resolution. \textbf{\textit{Right panel:}} Scattered light 2\,$\mu$m polarized fraction map with polarization vectors superimposed.}
        \label{fig:model_1_zone}
    \end{figure}
\end{center}

In Figure \ref{fig:model_1_zone}, we present the same model as PDS 453 but with only one zone defined between $R_{in} = 0.2$\,au and $R_{out} = 160$\,au. The dust mass is $M_d = 1.8\times10^{-5}\,M_\odot$, the scale height is fixed at $H_0=8.5$\,au, the flaring index at $\beta=1.13$ and the surface density at $p=-1$.

\end{appendix}

\label{LastPage}
\end{document}